\documentclass[twocolumn, journal]{IEEEtran}
\IEEEoverridecommandlockouts

\usepackage{color}
\usepackage{xcolor}
\usepackage{bm}
\usepackage{graphicx}
\usepackage{amsmath}
\usepackage{amssymb}
\usepackage{multirow}
\usepackage{booktabs}
\usepackage{array}
\usepackage{amsthm}
\usepackage{lipsum}
\usepackage{enumerate}
\usepackage{stfloats}
\usepackage{subfigure}
\usepackage{cases}
\usepackage{cite}

\allowdisplaybreaks

\title{Beyond Diagonal Reconfigurable Intelligent Surfaces in Wideband OFDM Communications: Circuit-Based Modeling and Optimization}

\author{Hongyu Li,~\IEEEmembership{Member,~IEEE}, Matteo Nerini,~\IEEEmembership{Member,~IEEE}, \\Shanpu Shen,~\IEEEmembership{Senior Member,~IEEE}, and Bruno Clerckx,~\IEEEmembership{Fellow,~IEEE}
\thanks{Manuscript received 12 May 2024; revised 22 October 2024 and 14 January 2025; accepted 15 January 2025. 
The associate editor coordinating the review of this article
and approving it for publication was Prof. Hien Ngo.
This work has been partially supported by UKRI grant EP/Y004086/1, EP/X040569/1, EP/Y037197/1, EP/X04047X/1, EP/Y037243/1. 
Part of this work has been submitted to IEEE International Workshop on Signal Processing Advances in Wireless Communications (SPAWC), 2024 \cite{li2024wideband}.}
\thanks{Hongyu Li is with the Internet of Things Thrust, The Hong Kong University of Science and Technology (Guangzhou), Guangzhou, Guangdong 511400, China (e-mail: hongyuli@hkust-gz.edu.cn).}
\thanks{Matteo Nerini and Bruno Clerckx are with the Department of Electrical and Electronic Engineering, Imperial College London, London SW7 2AZ, U.K. (e-mail: \{m.nerini20,~b.clerckx\}@imperial.ac.uk).}
\thanks{Shanpu Shen is with the Department of Electrical Engineering and Electronics, University of Liverpool, Liverpool L69 3GJ, U.K. (e-mail: Shanpu.Shen@liverpool.ac.uk).}
}

\begin{document}
\maketitle
\thispagestyle{empty}
\begin{abstract}
    This work investigates the modeling and optimization of beyond diagonal reconfigurable intelligent surface (BD-RIS), which generalizes conventional RIS with diagonal phase shift matrices and provides additional flexibility for manipulating wireless channels, in wideband communication systems. Specifically, we start from the signal modeling of the BD-RIS-aided orthogonal frequency division multiplexing (OFDM) system, which bridges the time-domain and frequency-domain channels, and explicitly shows the frequency dependence of the BD-RIS response. We next characterize the frequency dependence of the BD-RIS response based on circuit models. Benefiting from the admittance parameter analysis, we model individually each tunable admittance component of BD-RIS and derive an approximated linear expression with respect to the frequency of the transmit signals. With the proposed signal model for the BD-RIS-aided OFDM system and the frequency-dependent BD-RIS model, we propose algorithms to optimize the BD-RIS and the power allocation at the transmitter to maximize the average rate for a BD-RIS-aided OFDM system. 
    % We study BD-RIS with continuous as well as discrete-value tunable admittance components, and different architectures.
    Finally, simulation results show that BD-RIS outperforms conventional RIS in the OFDM system. More importantly, the impact of wideband modeling of BD-RIS on the system performance becomes more significant as the circuit complexity of BD-RIS architectures increases. 
\end{abstract}


\maketitle

\section{Introduction}

Beyond diagonal reconfigurable intelligent surface (BD-RIS) has been recently proposed to break through the limitation of conventional RIS with diagonal phase shift matrices \cite{wu2021intelligent,di2020smart,gong2020toward,liu2021reconfigurable} and provide enhanced channel gain and enlarged coverage \cite{li2023reconfigurable}. 
The concept of BD-RIS has been first proposed in \cite{shen2021}, where group- and fully-connected architectures have been introduced by interconnecting part of/all the RIS elements, respectively, with additional tunable impedance components to enhance the channel strength compared to conventional RIS.
Following \cite{shen2021}, a global optimal solution to maximize the received power for BD-RIS-aided wireless systems has been proposed in \cite{nerini2023closed}. 
Other architectures, namely tree- and forest-connected architectures, have been proposed in \cite{nerini2023beyond} to reduce the circuit design complexity of BD-RIS while maintaining satisfactory performance.
Thanks to the flexible architectures and antenna arrangements, BD-RISs supporting hybrid transmitting and reflecting mode \cite{li2022beyond} and multi-sector mode \cite{li2023beyond} have been proposed, which generalize and go beyond the intelligent omni-surface (IOS) \cite{zhang2022intelligent} to achieve full-space coverage while providing enhanced performance.
Further, the channel estimation strategy suitable for BD-RISs with different architectures and modes has been proposed in \cite{li2024channel}, where BD-RIS still outperforms conventional RIS in the presence of channel estimation error.
BD-RIS has also been theoretically proven to enable the implementation of optimal and low complexity stacked intelligent metasurfaces (SIM) \cite{an2023stacked,nerini2024physically}.
The aforementioned works \cite{li2023reconfigurable,shen2021,nerini2023closed,nerini2023beyond,li2022beyond,li2023beyond,li2024channel,nerini2024physically} mainly focus on using the scattering parameter analysis\footnote{In microwave engineering, multi-port network analysis is a powerful and useful technique for analyzing wireless systems, by regarding each antenna in the wireless system as a port and characterizing the behavior between its terminal voltage, current, incident wave, and reflected wave \cite{pozar2011microwave}. Specifically, the scattering parameter of a multi-port network relates the voltage waves incident on the ports to those reflected from the ports.} to model, analyze, and design the BD-RIS-aided communication system.
On top of this, a universal framework has been proposed in \cite{nerini2023universal} to unify the scattering, impedance, and admittance parameters analysis for characterizing the RIS-aided communication models.
Insightful discussions about the preferable parameter analyses for different use cases have also been provided. 
Exploiting the advantages of admittance parameter analysis in characterizing the circuit model of BD-RIS \cite{nerini2023universal}, the practical loss due to the inter-element connections has been investigated based on the transmission line theory \cite{nerini2024localized}, which again demonstrates the enhanced channel gain provided by BD-RIS over conventional RIS in the presence of losses. 

Although the advantages of BD-RIS have been shown from multiple aspects \cite{li2023reconfigurable,shen2021,nerini2023closed,nerini2023beyond,li2022beyond,li2023beyond,li2024channel,nerini2023universal,nerini2024localized,nerini2024physically}, it is worth noticing that the majority of existing BD-RIS works only consider narrowband scenarios. 
Two recent works \cite{soleymani2024maximizing,demir2024wideband} have studied the beamforming design of BD-RIS in wideband scenarios. 
However, the limitation of \cite{soleymani2024maximizing,demir2024wideband} is that they assume the BD-RIS model is frequency-independent, which may be inaccurate for wideband applications because the scattering matrix of BD-RIS determined by the reconfigurable impedance network is essentially frequency-dependent, as already highlighted in the conventional RIS literature \cite{wang2024wideband,hu2023wideband}.
Therefore, how to accurately model the wideband BD-RIS response and explore the benefits of BD-RIS in wideband communication systems remain important open problems.

While there are a few works on the wideband modeling and optimization of conventional RIS \cite{li2021intelligent,jiang2021general,zhang2021joint,zheng2019intelligent,bai2021resource}, the generalization to BD-RIS is difficult due to the following twofold challenges.
\textit{First}, from the modeling perspective, in conventional RIS, each element is not connected to each other such that the phase and amplitude response for each element can be independently modelled based on specific circuit designs \cite{li2021intelligent}. Nevertheless, this does not hold for BD-RIS with interconnected elements, since the phase and amplitude of each entry of the scattering matrix also depend on the other entries, which makes it impossible to individually model each entry of the scattering matrix of BD-RIS architectures.
\textit{Second}, from the optimization perspective, the scattering matrix of BD-RIS has more intricate constraints than conventional RIS arising from both the frequency dependence in the wideband model and the beyond diagonal property, which means that existing conventional RIS optimization methods for wideband scenarios \cite{jiang2021general,zhang2021joint,zheng2019intelligent,bai2021resource} and BD-RIS optimization methods for narrowband scenarios \cite{nerini2023beyond,nerini2023closed,shen2021,li2023reconfigurable} cannot be applied in wideband BD-RIS scenarios.

To address the above two challenges of wideband BD-RIS, in this work, we model and optimize the wideband BD-RIS by leveraging the advantages of admittance parameter analysis. This leads to the following contributions.

\textit{First}, we derive the signal model for the BD-RIS aided single input single output orthogonal frequency division multiplexing (SISO-OFDM) system. 
Specifically, we start from the time-domain channels between the transmitter to the receiver, and then bridge them to the frequency-domain forms to provide a comprehensive understanding of the behavior of BD-RIS in both domains. Different from the derivation in \cite{demir2024wideband} where the frequency dependence of the BD-RIS response is ignored, the derived signal model in this paper explicitly shows the frequency dependence of the BD-RIS response in wideband communication systems. 
Meanwhile, different from the derivation in \cite{li2021intelligent} where the conventional RIS is directly modeled in the frequency domain while its time-domain model remains unknown, the derived model characterizes BD-RIS in the time domain. This allows us to explain BD-RIS from the time domain: the interconnections construct multi-tap impulse response between BD-RIS elements.

\textit{Second}, we propose a novel wideband model of BD-RIS based on lumped circuit models. Specifically, we express BD-RIS using admittance parameter analysis, such that the frequency dependence of each tunable admittance component can be individually characterized. 
The proposed model has simple linear expressions while capturing accurately the frequency dependence of the BD-RIS admittance matrix within practical frequency ranges. 
The proposed model is also sufficiently general to adapt to any BD-RIS architecture, such as group- and forest-connected architectures.

\textit{Third}, based on the proposed wideband BD-RIS model, we optimize the admittance matrix of BD-RIS with group- and forest-connected architectures to maximize the average rate for BD-RIS-aided SISO-OFDM when each tunable admittance has either continuous or discrete values. 
Specifically, for the continuous-value case, we propose to transform the original problem into the unconstrained optimization, which is directly solved by the well-known quasi-Newton method.
For the discrete-value case, we adopt a greedy algorithm to iteratively optimize the tunable admittance components of BD-RIS. 

\textit{Fourth}, we evaluate the performance of the proposed optimization methods through numerical simulations. 
Results show that BD-RIS always outperforms conventional RIS whether the wideband modeling of RIS is taken into account in the optimization process or not.
In addition, BD-RIS with the proposed wideband modeling and optimization outperforms that with the frequency-independent model \cite{demir2024wideband}.
Furthermore, the impact of accurately modeling BD-RIS in wideband scenarios becomes more significant for BD-RIS with more tunable admittance components in the architecture.
That is, when using conventional RIS and simple BD-RIS architectures such as group/forest-connected with group size no more than 3, it is reasonable to directly adopt the frequency-independent BD-RIS model; for more advanced architectures such as group-connected with group size 6, proper wideband modeling is advised.

\textit{Organization:} 
Section \ref{sc:System} derives the BD-RIS-aided SISO-OFDM communication model. Section \ref{sc:Modeling} introduces the circuit-based wideband BD-RIS model using admittance parameter analysis. Section \ref{sc:Optimization_SISO} applies the wideband BD-RIS model and optimizes the beamforming for BD-RIS-aided SISO-OFDM systems.  
Section \ref{sc:Performance} evaluates the performance of the proposed wideband BD-RIS model and optimization methods. 
Section \ref{sc:Conclusion} concludes this work.

\textit{Notations:}
Boldface lower-case and upper-case letters indicate vectors and matrices, respectively. 
$\mathbb{C}$ and $\mathbb{R}$ denote the set of complex and real numbers, respectively.
$\mathbb{E}\{\cdot\}$ denotes the statistical expectation.
$(\cdot)^T$, $(\cdot)^H$, and $(\cdot)^{-1}$ denote the transpose, conjugate-transpose, and inverse, respectively.
$\Re\{\cdot\}$ and $\Im\{\cdot\}$ denote the real and imaginary parts of a complex number, respectively.
$\mathsf{diag}(a_1,\ldots,a_N)$ and $\mathsf{blkdiag}(\mathbf{A}_1,\ldots,\mathbf{A}_N)$ denote diagonal and block-diagonal matrices, respectively, with diagonal entries $a_1,\ldots,a_N$ and diagonal blocks $\mathbf{A}_1,\ldots,\mathbf{A}_N$.
$\mathsf{vec}(\cdot)$ and $\overline{\mathsf{vec}}(\cdot)$ denote the vectorization and its reverse operation, respectively.
$|\cdot|$ denotes the absolute-value norm of a scalar or the size of a set. 
$\jmath = \sqrt{-1}$ denotes the imaginary unit.
$\mathbf{I}_N$ denotes the $N\times N$ identical matrix. 
$\mathbf{0}_{M\times N}$ denotes the $M\times N$ all-zero matrix. 
$[\mathbf{A}]_{i,j}$ and $[\mathbf{a}]_i$, respectively, denote the $(i,j)$-th entry of $\mathbf{A}$ and the $i$-th entry of $\mathbf{a}$. 
$[\mathbf{a}]_{i:j}$ extracts the $i$-th through the $j$-th entries of $\mathbf{a}$.
$\mathbf{a}\sim\mathcal{CN}(\mathbf{0}_{N\times 1},\sigma^2\mathbf{I}_N)$ characterizes a column vector $\mathbf{a}$ with entries following the circular symmetric complex Gaussian (CSCG) distribution. 
$\mathbf{F}_N\in\mathbb{C}^{N\times N}$ denotes the normalized discrete Fourier transform (DFT) matrix with $[\mathbf{F}_N]_{i,j} = \frac{1}{\sqrt{N}}\exp(-\jmath 2\pi\frac{(i-1)(j-1)}{N})$. 
A block-circulant matrix $\mathbf{C}\in\mathbb{C}^{M_1N\times M_2N}$ is defined based on a sequence of blocks $\mathbf{C}_0\in\mathbb{C}^{M_1\times M_2}, \ldots, \mathbf{C}_{N-1}\in\mathbb{C}^{M_1\times M_2}$,
such that all columns are composed of the same blocks and each $M_2$ columns are rotated one block to the bottom relative to the preceding $M_2$ columns. That is, $\mathbf{C}$ takes the form 
\begin{equation}
    \label{eq:circulant_mtx}
    \begin{aligned}
        \mathbf{C} &= \mathsf{blkcirc}(\mathbf{C}_0,\mathbf{C}_1,\ldots,\mathbf{C}_{N-1})\\ 
        &= \begin{bmatrix}
            \mathbf{C}_0 & \mathbf{C}_{N-1} & \cdots &\mathbf{C}_{2} &\mathbf{C}_1\\
            \mathbf{C}_1 & \mathbf{C}_0 & \mathbf{C}_{N-1} & ~ & \mathbf{C}_{2}\\
            \vdots  & \mathbf{C}_{1} & \mathbf{C}_{0} &\ddots &\vdots\\
            \mathbf{C}_{N-2} &~ &\ddots &\ddots &\mathbf{C}_{N-1}\\
            \mathbf{C}_{N-1} &\mathbf{C}_{N-2} &\cdots &\mathbf{C}_{1} &\mathbf{C}_{0}
        \end{bmatrix}.
    \end{aligned}
\end{equation}
The circulant matrix is a special case of the block-circulant matrix with $M_1=M_2=1$.

\section{System Model for BD-RIS-Aided SISO-OFDM}
\label{sc:System}

\begin{figure*}
    \centering
    \includegraphics[width=0.85\textwidth]{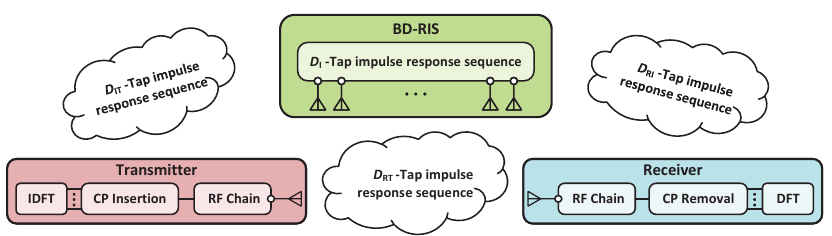}
    \caption{Diagram for the BD-RIS-aided SISO-OFDM system.}\label{fig:syst_mod}
\end{figure*}

We consider an $M$-element BD-RIS-aided wideband SISO-OFDM system with $N$ subcarriers\footnote{In this work, we focus on BD-RIS-aided SISO-OFDM systems for the ease of illustration, while the signal modeling in this work can be readily extended to multiple input multiple output (MIMO) OFDM systems.}. 
The signal model for the BD-RIS-aided SISO-OFDM system is explained below and illustrated in Fig. \ref{fig:syst_mod}. 

At the transmitter, we denote $s_n\in\mathbb{C}$ the transmit symbol associated with subcarrier $n$, $\mathbb{E}\{|s_n|^2\} = 1$, $\forall n\in\mathcal{N}= \{1,\ldots,N\}$. 
The transmitter allocates power $p_n\ge 0$ to symbol $s_n$, and has a total power budget $P$ such that $\sum_{n\in\mathcal{N}}p_n\le P$.
This leads to the discrete-time-domain signal $\bar{\mathbf{s}}\in\mathbb{C}^{N\times 1}$ as
\begin{equation}
    \bar{\mathbf{s}}=\mathbf{F}_N^H\mathbf{s},
\end{equation} 
where $\mathbf{s} = [\sqrt{p_1}s_1,\ldots,\sqrt{p_N}s_N]^T\in\mathbb{C}^{N\times 1}$ is the frequency-domain transmit symbol vector.
After adding the cyclic prefix (CP), the time-domain signal is up-converted to the radio frequency (RF) domain by the RF chain and radiated to the wireless environment through the transmit antenna \cite{goldsmith2005wireless,stuber2004broadband}.

In the BD-RIS-aided SISO-OFDM system, we define the channel from the transmit antenna to the receive antenna as a $D_{RT}$-tap finite-duration impulse response sequence $\{\bar{h}_{RT,i}\in\mathbb{C}~|~\forall i = 0, \ldots,D_{RT}-1\}$.
In addition, the channel from the transmit antenna to BD-RIS is modeled as a $D_{IT}$-tap impulse response sequence $\{\bar{\mathbf{h}}_{IT,i}\in\mathbb{C}^{M\times 1}~|~ \forall i = 0, \ldots,D_{IT}-1\}$. The channel from the BD-RIS to the receive antenna is modeled as a $D_{RI}$-tap impulse response sequence $\{\bar{\mathbf{h}}_{RI,i}\in\mathbb{C}^{1\times M}~|~\forall i = 0,\ldots,D_{RI}-1\}$. 
The response of BD-RIS can be regarded as the channel between the BD-RIS elements and themselves, which is manipulated by the BD-RIS tunable circuit.
In contrast to conventional RIS where each element is not connected with each other, in BD-RIS, different elements are connected to each other by additional tunable impedance components \cite{shen2021} such that multiple paths between elements exist. 
More details about the circuit model of BD-RIS can be found in \cite{shen2021,nerini2023beyond}, and will be briefly explained in Section \ref{sc:Modeling}-A.
The fact that those connections generate multiple paths between BD-RIS elements motivates us to describe BD-RIS as a $D_I$-tap impulse response sequence $\{\bar{\mathbf{\Theta}}_i\in\mathbb{C}^{M\times M}~|~\forall i = 0,\ldots,D_I-1\}$ with 
$\bar{\mathbf{\Theta}}_i$ not being limited to be diagonal thanks to the inter-element connections. 
As such, the time-domain channel from the transmitter through BD-RIS to the receiver is the series convolution between $\bar{\mathbf{h}}_{RI,i}$, $\bar{\mathbf{\Theta}}_i$, and $\bar{\mathbf{h}}_{IT,i}$.

At the receiver, the discrete-time-domain received signal after down-converting to the baseband and removing the CP is the convolution of the time-domain channels and the time-domain symbols \cite{goldsmith2005wireless,stuber2004broadband}.
Since the circular convolution of discrete sequences can be expressed as a matrix multiplication by introducing (block) circulant matrices, we write the discrete-time-domain received signal $\bar{\mathbf{y}}\in\mathbb{C}^{N\times 1}$ in the following compact form
\begin{equation}
    \bar{\mathbf{y}} = (\bar{\mathbf{H}}_{RT} + \bar{\mathbf{H}}_{RI}\bar{\mathbf{\Theta}}\bar{\mathbf{H}}_{IT})\bar{\mathbf{s}} + \bar{\mathbf{z}}, 
\end{equation}
where $\bar{\mathbf{z}}\thicksim\mathcal{CN}(\mathbf{0}_{N\times 1},\sigma^2\mathbf{I}_N)$ is the noise. 
$\bar{\mathbf{H}}_{RT}\in\mathbb{C}^{N\times N}$, $\bar{\mathbf{H}}_{RI}\in\mathbb{C}^{N\times NM}$, $\bar{\mathbf{\Theta}}\in\mathbb{C}^{NM\times NM}$, and $\bar{\mathbf{H}}_{IT}\in\mathbb{C}^{NM\times N}$ are (block) circulant matrices constructed as in (\ref{eq:circulant_mtx}), based on respectively the following sequence entries/blocks:
\begin{equation}
    \begin{aligned}
        &\bar{\mathbf{H}}_{RT} = \mathsf{blkcirc}(\bar{h}_{RT,0},\ldots,\bar{h}_{RT,D_{RT}-1}, \underbrace{0,\ldots,0}_{N-D_{RT}~\text{copies}}),\\
        &\bar{\mathbf{H}}_{RI} = \mathsf{blkcirc}(\bar{\mathbf{h}}_{RI,0},\ldots,\bar{\mathbf{h}}_{RI,D_{RI}-1}, \underbrace{\mathbf{0}_{1\times M},\ldots,\mathbf{0}_{1\times M}}_{N-D_{RI}~\text{copies}}),\\
        &\bar{\mathbf{\Theta}} = \mathsf{blkcirc}(\bar{\mathbf{\Theta}}_{0},\ldots,\bar{\mathbf{\Theta}}_{D_{I}-1}, \underbrace{\mathbf{0}_{M\times M},\ldots,\mathbf{0}_{M\times M}}_{N-D_{I}~\text{copies}}),\\
        &\bar{\mathbf{H}}_{IT} = \mathsf{blkcirc}(\bar{\mathbf{h}}_{IT,0},\ldots,\bar{\mathbf{h}}_{IT,D_{IT}-1}, \underbrace{\mathbf{0}_{M\times 1},\ldots,\mathbf{0}_{M\times 1}}_{N-D_{IT}~\text{copies}}).        
    \end{aligned}
\end{equation}
After operating DFT, the received signal $\mathbf{y}\in\mathbb{C}^{N\times 1}$ in the frequency-domain is given by 
\begin{equation}
    \mathbf{y} = \mathbf{F}_N\bar{\mathbf{y}} =  \underbrace{\mathbf{F}_N(\bar{\mathbf{H}}_{RT} + \bar{\mathbf{H}}_{RI}\bar{\mathbf{\Theta}}\bar{\mathbf{H}}_{IT})\mathbf{F}_N^H}_{=\mathbf{H}\in\mathbb{C}^{N\times N}}\mathbf{s} + \underbrace{\mathbf{F}_N\bar{\mathbf{z}}}_{=\mathbf{z}\in\mathbb{C}^{N\times 1}},
\end{equation}
with the frequency-domain SISO-OFDM channel $\mathbf{H}$ and the noise $\mathbf{z}\sim \mathcal{CN}(\mathbf{0}_{N\times 1},\sigma^2\mathbf{I}_N)$.
In the following proposition, we show that the frequency-domain channel $\mathbf{H}$ can be equivalently described by the frequency-domain channels from the transmitter to the receiver, from the transmitter to the BD-RIS, and from the BD-RIS to the receiver, as well as the frequency-dependent scattering matrix of BD-RIS.

\textit{Proposition 1:} 
The frequency-domain BD-RIS-aided SISO-OFDM channel $\mathbf{H}$ is a diagonal matrix, i.e., $\mathbf{H} = \mathsf{diag}(h_1,\dots,h_N)$. Specifically, each diagonal entry $h_n\in\mathbb{C}$ denotes the BD-RIS-aided channel at subcarrier $n$, $\forall n\in\mathcal{N}$ and is given by 
\begin{equation}
    \label{eq:channel_s}
    h_n = h_{RT,n} + \mathbf{h}_{RI,n}\mathbf{\Theta}_n\mathbf{h}_{IT,n}, \forall n,
\end{equation}
where $h_{RT,n}\in\mathbb{C}$, $\mathbf{h}_{RI,n}\in\mathbb{C}^{1\times M}$, and $\mathbf{h}_{IT,n}\in\mathbb{C}^{M\times 1}$, respectively, are the frequency-domain channels from the transmitter to the receiver, from the BD-RIS to the receiver, and from the transmitter to the BD-RIS, and $\mathbf{\Theta}_n\in\mathbb{C}^{M\times M}$ is the scattering matrix of BD-RIS at subcarrier $n$. These frequency-domain channels are obtained from the time-domain ones based on the following mapping: 
\begin{equation}
    \begin{aligned} 
        &h_{RT,n} = [\mathbf{F}_N\bar{\mathbf{H}}_{RT}\mathbf{F}_N^H]_{n,n}, \forall n, \\
        &[\mathbf{h}_{RI,n}]_i = [\mathbf{F}_N\bar{\mathbf{H}}_{RI,i}\mathbf{F}_N^H]_{n,n},\forall i\in\mathcal{M},\forall n,\\
        &[\mathbf{h}_{IT,n}]_j = [\mathbf{F}_N\bar{\mathbf{H}}_{IT,j}\mathbf{F}_N^H]_{n,n},\forall j\in\mathcal{M},\forall n,\\
        &[\mathbf{\Theta}_{n}]_{i,j} = [\mathbf{F}_N\bar{\mathbf{\Theta}}_{i,j}\mathbf{F}_N^H]_{n,n},\forall i,j,n,
    \end{aligned}
\end{equation}
where $\mathcal{M}=\{1,\ldots,M\}$. $\bar{\mathbf{H}}_{RI,i}\in\mathbb{C}^{N\times N}$, $\bar{\mathbf{H}}_{IT,j}\in\mathbb{C}^{N\times N}$, and $\bar{\mathbf{\Theta}}_{i,j}\in\mathbb{C}^{N\times N}$ are circulant matrices extracted respectively from  $\bar{\mathbf{H}}_{RI}$, $\bar{\mathbf{H}}_{IT}$, and $\bar{\mathbf{\Theta}}$, with 
$[\bar{\mathbf{H}}_{RI,i}]_{:,p} = [\bar{\mathbf{H}}_{RI}]_{:,(p-1)M+i}$, $[\bar{\mathbf{H}}_{IT,j}]_{q,:} = [\bar{\mathbf{H}}_{IT}]_{(q-1)M+j,:}$, $[\bar{\mathbf{\Theta}}_{i,j}]_{p,q} =  [\bar{\mathbf{\Theta}}]_{(p-1)M+i,(q-1)M+j}$, 
$\forall i,j\in\mathcal{M},\forall p,q\in\mathcal{N}$.

\textit{Proof:} Please refer to the Appendix. \hfill$\square$ 

From Proposition 1, we observe that
\begin{enumerate}[$\bullet$]
    \item the obtained BD-RIS-aided wideband channel aligns with the intuition in conventional OFDM systems, i.e., the wideband channels in OFDM systems can be described as parallel narrowband channels at different subcarriers;
    \item the obtained channel has the same expression as the conventional RIS-aided wideband channel \cite{li2021intelligent}, except that the BD-RIS response is not limited to being diagonal;
    \item the scattering matrix of BD-RIS is frequency-dependent, i.e., in general $\mathbf{\Theta}_n\ne\mathbf{\Theta}_m$ for different subcarriers $n\ne m$.
\end{enumerate}

\textit{Remark 1:} It is worth noting that in the recently proposed adjustable-delay RIS \cite{an2024adjustable}, the frequency-dependent response of RIS is also captured in wideband OFDM systems. More specifically, the mathematical modeling of this kind of RIS is split into a delay matrix and a reflection coefficient (scattering) matrix. The frequency dependence of RIS response is reflected in the delay matrix, while the scattering matrix is assumed to be frequency independent, which is different from this work where the frequency dependence of the scattering matrix is explicitly studied.

\textit{Remark 2:} The derived model in (\ref{eq:channel_s}) is not merely a straightforward extension of the narrowband channel models widely used in existing literature \cite{wu2021intelligent,di2020smart,gong2020toward}, since the frequency dependence of BD-RIS response is explicitly captured. Meanwhile, instead of directly modeling RIS from the frequency domain as in \cite{swindlehurst2022channel,yue2023ris}, we model the BD-RIS from the time domain, i.e., $\bar{\mathbf{\Theta}}_i$, and link it to the frequency domain, i.e., $\mathbf{\Theta}_n$, to better understand the behavior of BD-RIS at two domains.

It is worth noting that in conventional RIS with diagonal phase shift matrix $\mathbf{\Theta}_n$, it is possible to individually model the phase and amplitude response of each diagonal entry of $\mathbf{\Theta}_n$ \cite{li2021intelligent}.
However, this method does not work for BD-RIS with scattering matrices $\mathbf{\Theta}_n$, $\forall n\in\mathcal{N}$, not limited to being diagonal, since their entries are highly coupled with each other due to the inter-element connections.
To address this challenge, we instead model the frequency-dependent admittance matrix of BD-RIS \cite{nerini2023universal}, where each entry of the admittance matrix can be expressed as the linear combination of tunable admittance components with each being independently controlled by circuit models.
As such, the modeling of the scattering matrix with coupled entries boils down to that of each tunable admittance component.
To that end, we map the scattering matrix $\mathbf{\Theta}_n$ to the admittance matrix $\mathbf{Y}_n\in\mathbb{C}^{M\times M}$, which directly characterizes the circuit model of BD-RIS, by
\begin{equation}
    \mathbf{\Theta}_n = (Y_0\mathbf{I}_M + \mathbf{Y}_n)^{-1}(Y_0\mathbf{I}_M - \mathbf{Y}_n), \forall n,
\end{equation}
where $Y_0$ refers to the characteristic admittance, e.g. $Y_0 = \frac{1}{50}$ siemens (S). 
The modelling of $\mathbf{Y}_n$, $\forall n\in\mathcal{N}$ will be elaborated based on the circuit model of BD-RIS as follows.

\section{Circuit-Based Wideband Modeling of BD-RIS}
\label{sc:Modeling}

In this section, we first review the BD-RIS modeling and architecture design in narrowband scenarios using admittance parameter analysis \cite{shen2021,nerini2023beyond}. Then, we, for the first time, establish the wideband BD-RIS model by characterizing its frequency dependence based on the circuit model.

\subsection{Admittance Matrix of BD-RIS}

Consider an $M$-element BD-RIS modeled as $M$ antennas connected to an $M$-port reconfigurable admittance network \cite{shen2021}. 
There are various architectures of the reconfigurable admittance network determined by the circuit topology design. In this work, we focus on the wideband modeling and optimization of two representative architectures, namely group-connected \cite{shen2021} and forest-connected \cite{nerini2023beyond}, which, respectively, include the fully- and tree-connected architectures as special cases. The mathematical descriptions are explained below. 

\subsubsection{Group-Connected Reconfigurable Admittance Network} 
\begin{figure}
    \centering
    \includegraphics[width=0.47\textwidth]{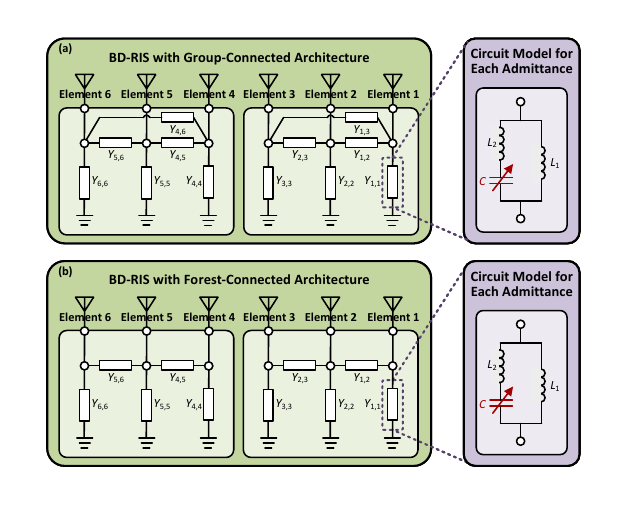}
    \caption{Examples of a 6-element BD-RIS with (a) group-connected and (b) forest-connected reconfigurable admittance network of group size 3, and the equivalent circuit for each admittance.}\label{fig:bdris_cir}
\end{figure}

For a general group-connected reconfigurable admittance network with $G$ uniform groups, the $\bar{M} = \frac{M}{G}$ ports in each group are all connected to each other by tunable admittance components, while ports in different groups are not connected to each other \cite{shen2021}. 
To facilitate understanding, an illustrative example of a 6-element BD-RIS with the group-connected architecture (of group size 3) is illustrated in Fig. \ref{fig:bdris_cir}(a).  
According to the circuit topology of the group-connected reconfigurable admittance network, the corresponding admittance matrix $\mathbf{Y}\in\mathbb{C}^{M\times M}$ has a block-diagonal structure defined as 
\begin{equation}
    \mathbf{Y} = \mathsf{blkdiag}(\bar{\mathbf{Y}}_1,\ldots,\bar{\mathbf{Y}}_G),
    \label{eq:Y_blkdiag}
\end{equation}
where $\bar{\mathbf{Y}}_g\in\mathbb{C}^{\bar{M}\times\bar{M}}$, $\forall g\in\mathcal{G} = \{1,\ldots,G\}$ is symmetric and purely imaginary for reciprocal and lossless reconfigurable admittance networks\footnote{In microwave engineering \cite{pozar2011microwave}, a reciprocal network means that the current at one port due to a voltage at a second port is the same as the current at the second port due to the same voltage at the first. This mathematically makes the admittance matrix of the network symmetric. A lossless network means the net power delivered to the network is zero, which mathematically makes the admittance matrix purely imaginary.}, that is 
\begin{equation}
    \bar{\mathbf{Y}}_g = \bar{\mathbf{Y}}_g^T, ~\Re\{\bar{\mathbf{Y}}_g\} = \mathbf{0}_{\bar{M}\times\bar{M}}, \forall g.
    \label{eq:Y_symmetric}
\end{equation}
In group $g$ of the group-connected reconfigurable admittance network, each port $m_{g} = (g-1)\bar{M}+m$ is connected to port $m'_g = (g-1)\bar{M}+m'$, with a tunable admittance\footnote{$Y_{m_g,m_g}$ refers to the admittance which connects port $m_g$ to ground.} $Y_{m_g,m'_g}$, $\forall m,m'\in\bar{\mathcal{M}}=\{1,\ldots,\bar{M}\}$, which satisfies $Y_{m_g,m'_g} = Y_{m'_g,m_g}$ for $m'>m$. 
Accordingly, $\bar{\mathbf{Y}}_g, \forall g\in\mathcal{G}$ writes as \cite{pozar2011microwave}
\begin{equation}
    [\bar{\mathbf{Y}}_g]_{m,m'} = \begin{cases}
        -Y_{m_g,m'_g}, &m_g\ne m'_g \\
        \sum_{k=(g-1)\bar{M}+1}^{g\bar{M}}Y_{m_g,k}, & m_g = m'_g
    \end{cases}.
    \label{eq:Y_calculate}
\end{equation}

\subsubsection{Forest-Connected Reconfigurable Admittance Network} 
In BD-RIS, the hardware cost mainly comes from the circuit complexity, that is the required number of tunable admittance components.
The circuit complexity of group-connected architecture, that is the number of tunable admittance components, is $\frac{\bar{M}(\bar{M}+1)}{2}G$, which grows quadratically with the group size $\bar{M}$. 
This makes it difficult to implement group-connected architecture with larger group size, e.g., $\bar{M} > 4$. 
To simplify the circuit design and reduce the circuit complexity while maintaining the beam manipulation flexibility, the forest-connected architecture of BD-RIS is proposed in \cite{nerini2023beyond}.
For a forest-connected reconfigurable admittance network with $G$ uniform groups, every two adjacent ports of the $\bar{M}$ ports in each group are connected via one tunable admittance component, referred to as tree-connected, while ports from different groups are disconnected \cite{nerini2023beyond}. As such, the circuit complexity reduces to $(2\bar{M}-1)G$. To facilitate understanding, an example of a 6-element BD-RIS with the forest-connected architecture is illustrated in Fig. \ref{fig:bdris_cir}(b). In this case, the admittance matrix $\mathbf{Y}$ still satisfies (\ref{eq:Y_blkdiag})-(\ref{eq:Y_calculate}). 
The difference compared to the group-connected architecture is that in each group $g$, ports $m_g$ and $m_g'$ are not connected if $|m_g - m_g'|>1$, $\forall m,m'\in\bar{\mathcal{M}}$, $\forall g\in\mathcal{G}$.
This leads to the following additional constraint:
\begin{equation}
    Y_{m_g,m'_g} = 0, \forall m_g, m'_g,~|m_g-m'_g| > 1.
    \label{eq:Y_forest}
\end{equation}

\textit{Remark 3:}
Group-connected BD-RIS includes conventional RIS (single-connected with $\bar{M}=1$) and fully-connected BD-RIS ($\bar{M}=M$) as two special cases. 
Meanwhile, forest-connected BD-RIS includes conventional RIS ($\bar{M}=1$) and tree-connected BD-RIS ($\bar{M}=M$) as two special cases. 
In addition, group- and forest-connected BD-RISs with $\bar{M}=2$  share the same circuit topology. 

In equations (\ref{eq:Y_blkdiag})-(\ref{eq:Y_forest}), we have general expressions of the admittance matrix for BD-RIS without specifying its frequency dependence, which will be identified in the following subsection based on the circuit model of each tunable admittance.

\subsection{Wideband Modeling of Tunable Admittance}

\begin{figure}
    \centering
    \subfigure[Susceptance versus frequency]{
    \includegraphics[width=0.22\textwidth]{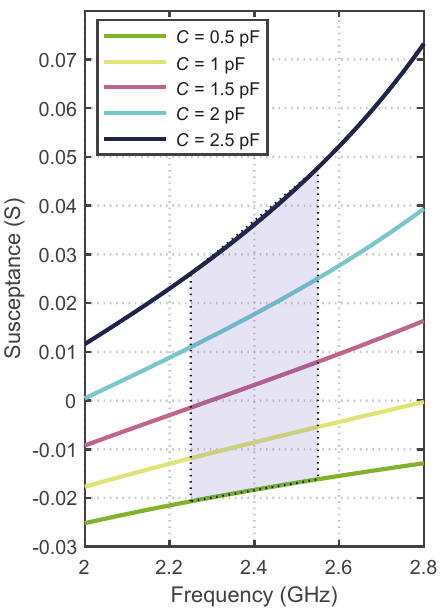}}
    \subfigure[Susceptance variation]{
        \includegraphics[width=0.22\textwidth]{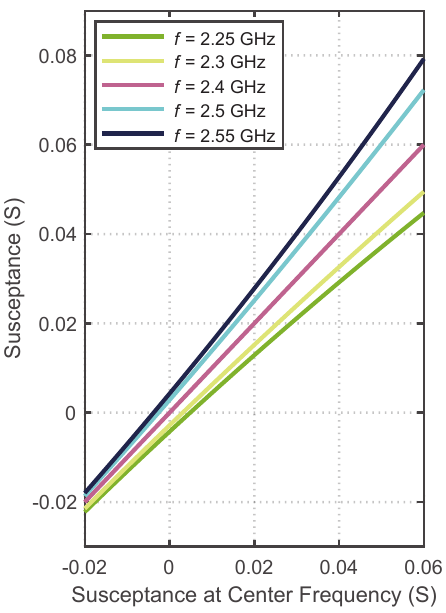}}
    \caption{The susceptance as a function of (a) frequency and (b) the value of susceptance at the center frequency $f_\mathrm{c} = 2.4$ GHz with $L_1 = 2.5$ nH, $L_2 = 0.7$ nH, and $C\in[0.2,3]$ pF for a practical varactor diode.}
    \label{fig:susceptance_f}
\end{figure}

In this work, we model each tunable admittance component as a lumped circuit consisting of two inductors $L_1$ and $L_2$, and one tunable capacitor $C$ \cite{koziel2013surrogate}, as illustrated in Fig. \ref{fig:bdris_cir}\footnote{Here we remove the equivalent resistance in the circuit to make the admittance network lossless. Detailed studies on the impact of lossy admittance networks can be found in \cite{nerini2024localized}.}. 
The admittance of the circuit is thus a function of the value of the capacitor and the frequency $f$ for the incident signals, which is given by\footnote{In this model, we set fixed and same values of inductors $L_1$ and $L_2$ for each admittance component, while making the value of the capacitor tunable and different, i.e., $C_{m_g,m'_g}$, for different admittance components.}
\begin{equation}
    Y_{m_g,m'_g} = \frac{1}{\jmath\omega L_1} + \frac{1}{\jmath \omega L_2 + \frac{1}{\jmath \omega C_{m_g,m'_g}}}, \forall m_g,m'_g,
    \label{eq:admittance_C_f}
\end{equation}
where $\omega=2\pi f$ denotes the angular frequency.
This yields the corresponding susceptance as 
$B_{m_g,m'_g} = \Im\{Y_{m_g,m'_g}\}$, $\forall m,m'\in\bar{\mathcal{M}}$, $\forall g\in\mathcal{G}$.
From (\ref{eq:admittance_C_f}) we can observe that the tunable admittance values at different frequencies are coupled to each other, that is for a certain value of capacitor $C$, the admittance will have different values at different frequencies. 
More importantly, this is an intrinsic phenomenon coming from the hardware implementation of BD-RIS elements, which cannot be simply ignored in BD-RIS-aided wideband communication systems. 
This phenomenon is also numerically illustrated in Fig. \ref{fig:susceptance_f}(a), where we consider a wideband system with center frequency $f_\mathrm{c} = 2.4$ GHz and adopt a practical varactor diode according to the data sheet for the diode SMV1231-079.  
Interestingly, the susceptance can be regarded as a linear function with respect to frequency for different values of $C_{m_g,m'_g}$ within some practical bandwidth for wideband communication systems, such as the range from 2.25 GHz to 2.55 GHz as highlighted in Fig. \ref{fig:susceptance_f}(a). 

\begin{figure}
    \centering
    \subfigure[Slope of equation (\ref{eq:B_Bc})]{
    \includegraphics[width=0.48\textwidth]{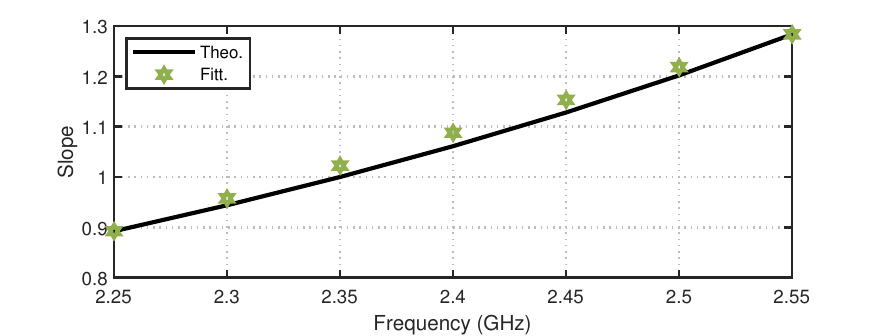}}
    \subfigure[Intercept of equation (\ref{eq:B_Bc})]{
    \includegraphics[width=0.48\textwidth]{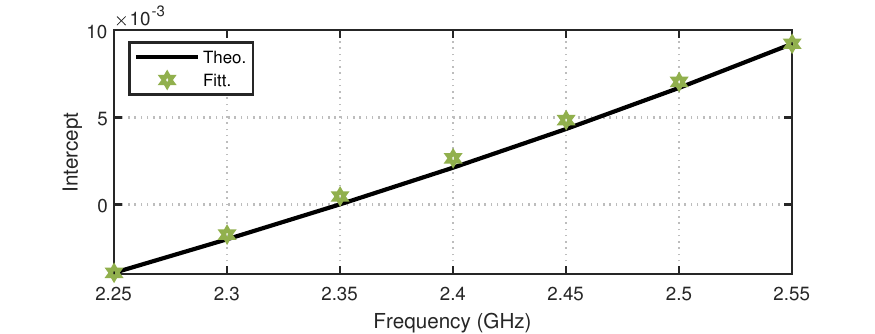}}
    \subfigure[Fitted model of equation (\ref{eq:B_Bc})]{
    \includegraphics[width=0.48\textwidth]{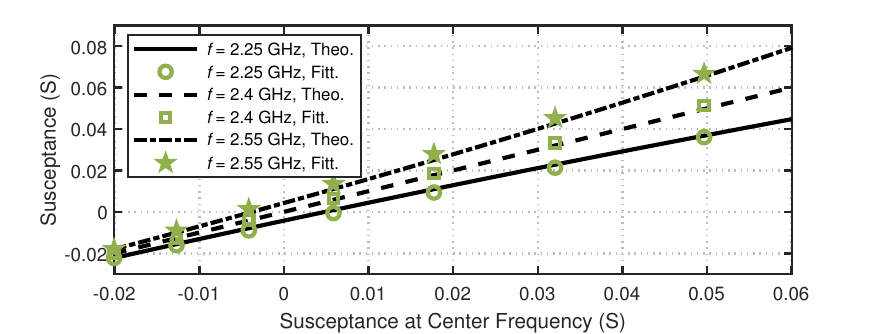}}
    \caption{(a) The slope and (b) the intercept of equation (\ref{eq:B_Bc}) as a function of $\omega$; (c) the fitted result based on equation (\ref{eq:fitted_model}). The parameters $\alpha_j$, $\beta_j$, $\forall j\in\{1,2\}$ in the linear model (\ref{eq:fitted_model}) are set as $\alpha_1 = 2.0046\times10^{-10}$, $\alpha_2 = -1.9968$, $\beta_1 = 6.2775\times10^{-12}$, $\beta_2 = -0.0942$.}
    \label{fig:fitted}
\end{figure}

To obtain more insights of how the value of susceptance varies in wideband systems, we define $B_{\mathrm{c},m_g,m'_g}$ as the susceptance at the center frequency $f_\mathrm{c}$. According to (\ref{eq:admittance_C_f}), we have the following relationship
\begin{equation}
    C_{m_g,m'_g} = \Bigg(\omega_\mathrm{c}^2 L_2 + \frac{\omega_\mathrm{c}}{B_{\mathrm{c},m_g,m'_g} + \frac{1}{\omega_\mathrm{c} L_1}}\Bigg)^{-1},
    \label{eq:admittance_c}
\end{equation}
where $\omega_\mathrm{c}=2\pi f_\mathrm{c}$.
Plugging (\ref{eq:admittance_c}) into (\ref{eq:admittance_C_f}), we can rewrite $B_{m_g,m'_g}$, $\forall m,m'\in\bar{\mathcal{M}}$, $\forall g\in\mathcal{G}$, as a function of $B_{\mathrm{c},m_g,m'_g}$ and $\omega$, that is 
\begin{equation}
    \begin{aligned}
        &B_{m_g,m'_g}(B_{\mathrm{c},m_g,m'_g},\omega)\\
         &= \frac{((\omega^2-\omega_\mathrm{c}^2)L_2+\omega^2L_1)\Big(B_{\mathrm{c},m_g,m'_g} + \frac{1}{\omega_\mathrm{c}L_1}\Big) - \omega_\mathrm{c}}{\omega L_1\Big(\omega_\mathrm{c} - (\omega^2-\omega_\mathrm{c}^2)L_2\Big(B_{\mathrm{c},m_g,m'_g} + \frac{1}{\omega_\mathrm{c}L_1}\Big)\Big)}.
    \end{aligned} \label{eq:B_Bc}
\end{equation}
It is difficult to characterize the relationship between $B_{m_g,m'_g}$ and $B_{\mathrm{c},m_g,m'_g}$ due to the complicated expression of (\ref{eq:B_Bc}). Fortunately, when focusing on the linear range highlighted in Fig. \ref{fig:susceptance_f}(a), we observe that $B_{m_g,m'_g}$ can be regarded as a linear function of $B_{\mathrm{c},m_g,m'_g}$, as illustrated in Fig. \ref{fig:susceptance_f}(b). 
Additionally, its slope and intercept can both be regarded as linear functions of $\omega$, which are shown in Figs. \ref{fig:fitted}(a) and \ref{fig:fitted}(b). 
The above observations motivate us to model the frequency dependence of each admittance component for BD-RIS in wideband communication systems as a simplified linear function, which is given by:
\begin{equation}\label{eq:fitted_model}
    \begin{aligned} 
        &B_{m_g,m'_g}(B_{\mathrm{c},m_g,m'_g},\omega) \\
        &~~~~~~~= F_1(\omega)B_{\mathrm{c},m_g,m'_g} + F_2(\omega), \forall m_g,m'_g,\\
        &F_1(\omega) = \alpha_1 \omega + \beta_1, ~~
        F_2(\omega) = \alpha_2 \omega + \beta_2,
    \end{aligned}
\end{equation} 
where parameters $\alpha_j$, $\beta_j$, $\forall j\in\{1,2\}$ are determined by $L_1$, $L_2$, and $\omega_\mathrm{c}$\footnote{In this work, we obtain the values of $\alpha_j$, $\beta_j$, $\forall j\in\{1,2\}$ by fitting the slope and intercept of (\ref{eq:B_Bc}) as linear functions of $\omega$ such that each of them can be characterized based on arbitrary two points on the lines.}. 
For the highlighted curves in Fig. \ref{fig:susceptance_f}(a), the theoretical results given by (\ref{eq:B_Bc}) and fitted results using the linear model (\ref{eq:fitted_model}) are shown in Fig. \ref{fig:fitted}. The results in Fig. \ref{fig:fitted}(c) guarantee a $0.27\%$ normalized mean square error between the fitted values and theoretical ones, which demonstrate the accuracy of the proposed linear model.

\textit{Remark 4:} The linear model (\ref{eq:fitted_model}) is a generalization of the narrowband model utilized in existing BD-RIS literature \cite{shen2021,nerini2023beyond,nerini2024localized}. Specifically, the narrowband scenario refers to the case that $\omega \approx \omega_\mathrm{c}$, yielding $F_1(\omega) \approx 1$ and $F_2(\omega) \approx 0$, such that the susceptance of each tunable admittance component is approximately frequency-independent.  

\subsection{BD-RIS with Frequency-Dependent Admittance Matrices}

In the BD-RIS-aided SISO-OFDM system, each sub-band $n$ has a carrier frequency $f_n$, within which we assume the admittance matrix of BD-RIS is constant in frequency, and only dependent on the carrier frequency $f_n$. 
Therefore, we characterize BD-RIS in the OFDM system by $N$ frequency-dependent admittance matrices $\mathbf{Y}_n$, $\forall n\in\mathcal{N}$, each of which 
refers to the admittance matrix of BD-RIS at subcarrier $n$. In addition, each $\mathbf{Y}_n$ is constructed based on equations (\ref{eq:Y_blkdiag})-(\ref{eq:Y_forest}) and the linear model (\ref{eq:fitted_model}) of each tunable admittance component. 
In the following, we will summarize the constraint of $\mathbf{Y}_n$ for BD-RIS with group- and forest-connected architectures. 

\subsubsection{Group-Connected BD-RIS}
Given the carrier frequency $f_n$ for subcarrier $n$, $\mathbf{Y}_n$ is constructed as 
\begin{subequations}\label{eq:Y_wideband_construction}
    \begin{align}
        \label{eq:blkdiag}
        &\mathbf{Y}_n = \mathsf{blkdiag}(\mathbf{Y}_{1,n},\ldots,\mathbf{Y}_{G,n}), \forall n, \\
        \label{eq:symmetric}
        &\mathbf{Y}_{g,n} = \jmath\mathbf{B}_{g,n}, ~\mathbf{B}_{g,n} = \mathbf{B}_{g,n}^T, \forall g,n,\\
        \label{eq:B_cal}
        &[\mathbf{B}_{g,n}]_{m,m'} = \begin{cases}
            -B_{m_g,m'_g,n}, &m_g \ne m'_g \\
            \sum_{k=(g-1)\bar{M}+1}^{g\bar{M}} B_{m_g,k,n}, & m_g = m'_g
        \end{cases},\\
        \label{eq:wideband_constraint}
        &B_{m_g,m'_g,n} = F_1(\omega_n)B_{\mathrm{c},m_g,m'_g} + F_2(\omega_n), \forall m_g,m'_g,n,\\
        \label{eq:range}
        &B_{\mathrm{c},m_g,m'_g} \in [B_\mathrm{min},B_\mathrm{max}], \forall m_g,m'_g,
    \end{align}
\end{subequations}
where $\omega_n = 2\pi f_n$, $B_\mathrm{min}$ and $B_\mathrm{max}$ are determined by the practical range of capacitor values. More specifically,
\begin{enumerate}[$\bullet$]
    \item when the capacitor has continuously tunable values, each $B_{\mathrm{c},m_g,m'_g}$ is arbitrarily selected within $[B_\mathrm{min},B_\mathrm{max}]$; 
    \item when the capacitor has finite tunable values, each $B_{\mathrm{c},m_g,m'_g}$ is selected from the set 
    \begin{equation}
        \mathcal{B} = \Big\{B_\mathrm{min} + \frac{B_\mathrm{max}-B_\mathrm{min}}{2^b -1}x|x=0,1,\ldots,2^b - 1\Big\},
    \end{equation} 
    where $b$ denotes the number of resolution bits. 
\end{enumerate}

\subsubsection{Forest-Connected BD-RIS}
Given the carrier frequency $f_n$ for subcarrier $n$, $\mathbf{Y}_n$ is constructed based on (\ref{eq:blkdiag})-(\ref{eq:B_cal}) and the following constraints: 
\begin{subequations}
    \begin{align}
        \label{eq:wideband_forest}
        &B_{m_g,m'_g,n} = \begin{cases}
            0, ~~\text{if~}|m_g-m'_g|> 1,&\\
            F_1(\omega_n)B_{\mathrm{c},m_g,m'_g}+ F_2(\omega_n), ~\text{otherwise},& 
        \end{cases}\\
        \label{eq:range_forest}
        &B_{\mathrm{c},m_g,m'_g} 
            \in[B_\mathrm{min},B_\mathrm{max}]~\text{or}~B_{\mathrm{c},m_g,m'_g}\in\mathcal{B}.
    \end{align} \label{eq:B_c_forest}
\end{subequations}

\textit{Remark 5:}
From (\ref{eq:Y_wideband_construction}) and (\ref{eq:B_c_forest}) we observe that $\mathbf{Y}_n$ for all carrier frequencies are dependent to each other and jointly tuned by $B_{\mathrm{c},m_g,m'_g}$. 
In \cite{soleymani2024maximizing,demir2024wideband}, however, the frequency-independent BD-RIS model is adopted in wideband systems. 
That is, the constraint (\ref{eq:wideband_constraint}) has always $F_1(\omega_n) = 1$ and $F_2(\omega_n) = 0$, such that $\mathbf{Y}_n = \mathbf{Y}$, $\forall n\in\mathcal{N}$. 
This simplified model facilitates the BD-RIS design but will cause performance loss due to the ignorance of frequency dependence of BD-RIS admittance matrices. The detailed performance comparison will be given in Section \ref{sc:Performance}. 

% In the following section, we will adopt the proposed BD-RIS model in the SISO-OFDM system and optimize the BD-RIS admittance matrices with both continuous and discrete values. 

\section{Optimization for BD-RIS-Aided SISO-OFDM}
\label{sc:Optimization_SISO}

In this section, we apply the proposed wideband BD-RIS model in a SISO-OFDM system to formulate the rate maximization problem and develop corresponding algorithms to optimize BD-RIS admittance matrices.

\subsection{Problem Formulation}

According to \cite{nerini2023universal}, by modeling the transmitter, BD-RIS, and receiver as a multi-port network, the wireless channel can be characterized by its admittance matrix, where the admittance matrix of BD-RIS, $\mathbf{Y}_n$, $\forall n\in\mathcal{N}$, is visible in the BD-RIS-aided wireless channel.
Specifically, the BD-RIS-aided wireless channel $h_n$ in (\ref{eq:channel_s}) can be re-written as \cite{nerini2023universal}
\begin{equation}
    h_n = \frac{1}{2Y_0}(-y_{RT,n} + \mathbf{y}_{RI,n}(\mathbf{Y}_n + Y_0\mathbf{I}_M)^{-1}\mathbf{y}_{IT,n}), \forall n,
    \label{eq:channel}
\end{equation}
where $y_{RT,n}\in\mathbb{C}$, $\mathbf{y}_{RI,n}\in\mathbb{C}^{1\times M}$, and $\mathbf{y}_{IT,n}\in\mathbb{C}^{M\times 1}$, respectively, refer to the channels from transmitter to receiver, from BD-RIS to receiver, and from transmitter to BD-RIS based on the admittance parameters \cite{nerini2023universal}. According to the discussions in \cite{nerini2023universal}, these channels are a characterization of the wireless channels equivalent to the widely used models based on the scattering parameters \cite{wu2021intelligent}. The detailed mapping between the channels based on the scattering parameters and those based on the admittance parameters can be found in \cite{nerini2023universal}, and is briefly summarized as\footnote{In this work, we assume the individual channels between devices are perfectly known at the transmitter to focus purely on the effect of wideband modeling at BD-RIS. However, the proposed designs also work when only the cascaded channel is known. In practice, the cascaded channel can be obtained by letting the transmitter or receiver consecutively transmit pilot signals and by changing the pattern at BD-RIS based on specific metrics \cite{swindlehurst2022channel}.
Due to the unique architectures of BD-RIS, the corresponding cascaded channel has different construction and dimension compared to conventional RIS, such that the conventional RIS pattern cannot be directly used to obtain the cascaded channel in BD-RIS systems \cite{li2024channel}.
Meanwhile, the individual channels can be obtained by equipping the BD-RIS with a few RF chains, such that conventional channel estimation schemes without RIS can be directly adopted.}
\begin{equation}
    \begin{aligned}
        &\mathbf{y}_{RI,n} = -2Y_0\mathbf{h}_{RI,n}, ~~ \mathbf{y}_{IT,n} = -2Y_0\mathbf{h}_{IT,n},\\
        &y_{RT,n} = -2Y_0(h_{RT,n} - \mathbf{h}_{RI,n}\mathbf{h}_{IT,n}), \forall n.
    \end{aligned}
\end{equation}

We aim to jointly design the power allocation at the transmitter and the susceptances $B_{\mathrm{c},m_g,m'_g}$ at the center frequency to maximize the average rate for the SISO-OFDM system. This yields the following problem:
\begin{equation}\label{eq:prob0}
    \begin{aligned}
    &\{B_{\mathrm{c},m_g,m'_g}^\star, p_n^\star\}_{\forall m_g,m'_g,n}\\
    &~~~= \arg~\max_{\substack{\text{(\ref{eq:Y_wideband_construction}) or (\ref{eq:blkdiag})-(\ref{eq:B_cal}),(\ref{eq:B_c_forest})},\\ \sum_n p_n\le P}}\frac{1}{N}\sum_{n\in\mathcal{N}}\log_2\Big(1+\frac{p_n|h_n|^2}{\sigma^2}\Big),
    \end{aligned}
\end{equation} 
where the constraints in (\ref{eq:Y_wideband_construction}) are considered for group-connected architectures while the constraints in (\ref{eq:blkdiag})-(\ref{eq:B_cal}) and (\ref{eq:B_c_forest}) for forest-connected architectures.
Problem (\ref{eq:prob0}) is a challenging joint optimization mainly due to the coupling constraints of BD-RIS in wideband systems. 
To simplify the design, we decouple problem (\ref{eq:prob0}) into two sub-problems and solve each of them individually. 

\subsection{Two-Stage Optimization}

The wideband modeling of BD-RIS couples the admittances at different subcarriers. These admittances are further embedded in the $\log(\cdot)$ function. Therefore, it is difficult to directly design the BD-RIS with the original objective function in problem (\ref{eq:prob0}). 
Meanwhile, the power allocation and the admittance matrices of BD-RIS are coupled, which further complicates the optimization. 
In addition, the objective function in (\ref{eq:prob0}) includes $N$ large-dimensional matrix inverses, each of which requires $\mathcal{O}(G\bar{M}^3)$ computational complexity. 
In this sense, jointly optimizing BD-RIS and power allocation will lead to unaffordable time consumption.
To simplify the problem and reduce the computational complexity, we decouple the design of BD-RIS and power allocation. This leads to the following two-stage optimization framework: 

\textit{Stage 1:}
We first simplify the objective function in (\ref{eq:prob0}) by using Jensen's inequality, which yields
$\frac{1}{N}\sum_{n\in\mathcal{N}}\log_2(1+\frac{p_n|h_n|^2}{\sigma^2}) \le \log_2(1+\frac{1}{N}\sum_{n\in\mathcal{N}}\frac{p_n|h_n|^2}{\sigma^2})$.
In addition, we remove the impact of power allocation when designing BD-RIS by setting $p_1 = \cdots = p_N$, while the power allocation will be designed after the admittance matrices of BD-RIS have been determined. Therefore, we solve the following sum channel gain maximization problem:
\begin{equation}
    \begin{aligned}
        &\{B_{\mathrm{c},m_g,m'_g}^\star\}_{\forall m_g,m'_g}
        = \arg\max_{\substack{\text{(\ref{eq:symmetric})-(\ref{eq:range}) or}\\ \text{(\ref{eq:symmetric}),(\ref{eq:B_cal}),(\ref{eq:B_c_forest})}}} \sum_{n\in\mathcal{N}}\Big|-y_{RT,n}\\
        &~~~~~~~~~+\sum_{g\in\mathcal{G}}\mathbf{y}_{RI,g,n}(\jmath\mathbf{B}_{g,n}+Y_0\mathbf{I}_{\bar{M}})^{-1}\mathbf{y}_{IT,g,n}\Big|^2,
    \end{aligned}\label{eq:sub_prob1}
\end{equation}
where $\mathbf{y}_{RI,g,n} = [\mathbf{y}_{RI,n}]_{(g-1)\bar{M}+1:g\bar{M}}$, and $\mathbf{y}_{IT,g,n} = [\mathbf{y}_{IT,n}]_{(g-1)\bar{M}+1:g\bar{M}}$, $\forall g\in\mathcal{G}$, $\forall n\in\mathcal{N}$. 

\textit{Stage 2:}
When $B_{\mathrm{c},m_g,m'_g}^\star$, $\forall m,m'\in\bar{\mathcal{M}}$, $\forall g\in\mathcal{G}$, are determined, the BD-RIS-aided wideband channels, $h_n^\star$, $\forall n$, can be obtained by (\ref{eq:Y_wideband_construction}), (\ref{eq:B_c_forest}), and (\ref{eq:channel}). Then, problem (\ref{eq:prob0}) boils down to the following conventional power allocation problem: 
\begin{equation}\label{eq:sub_prob_power}
    \begin{aligned}
    &\{p_n^\star\}_{\forall n}= \arg~\max_{\sum_n p_n\le P}\frac{1}{N}\sum_{n\in\mathcal{N}}\log_2\Big(1+\frac{p_n|h_n^\star|^2}{\sigma^2}\Big),
    \end{aligned}
\end{equation} 
which can be solved by the well-known water-filling method.

In the following subsections, we will elaborate on solving problem (\ref{eq:sub_prob1}) for both group- and forest-connected BD-RISs when each $B_{\mathrm{c},m_g,m'_g}$ has either continuous values or discrete values controlled by the set $\mathcal{B}$.

\subsection{Solution to Problem (\ref{eq:sub_prob1}): Continuous-Value Admittance}
We propose to rewrite (\ref{eq:sub_prob1}) into an unconstrained optimization solve it using the quasi-Newton method. The detailed mathematical transformations for group- and forest-connected architectures are explained below.

\subsubsection{Group-Connected BD-RIS}
We start by re-describing (\ref{eq:B_cal}) as a function which maps the matrix $\bar{\mathbf{B}}_{g,n}\in\mathbb{C}^{\bar{M}\times\bar{M}}$ with entries $[\bar{\mathbf{B}}_{g,n}]_{m,m'} = B_{m_g,m'_g,n}$ to the matrix $\mathbf{B}_{g,n}$, i.e., $\mathbf{B}_{g,n} = F_3(\bar{\mathbf{B}}_{g,n})$, where $F_3(\cdot)$ is given such that 
\begin{equation}
    [\mathbf{B}_{g,n}]_{m,m'} = \begin{cases}
        -[\bar{\mathbf{B}}_{g,n}]_{m,m'}, &m\ne m'\\
        \sum_{k\in\bar{\mathcal{M}}}[\bar{\mathbf{B}}_{g,n}]_{m,k},  &m=m'
    \end{cases}.
\end{equation}
In addition, the symmetric constraint of each $\mathbf{B}_{g,n}$ in (\ref{eq:symmetric}) and the linear mapping (\ref{eq:B_cal}) imply that each $\bar{\mathbf{B}}_{g,n}$ should be symmetric, i.e., $\bar{\mathbf{B}}_{g,n} = \bar{\mathbf{B}}_{g,n}^T$. 
In other words, each $\bar{\mathbf{B}}_{g,n}$ is essentially determined by its $\bar{M}$ diagonal and $\frac{\bar{M}(\bar{M}-1)}{2}$ lower-triangular (or upper-triangular) entries. 
This motivates us to rewrite $\bar{\mathbf{B}}_{g,n}$ as 
$\bar{\mathbf{B}}_{g,n} = \overline{\mathsf{vec}}(\mathbf{P}\mathbf{b}_{g,n})$, where
$\mathbf{b}_{g,n}\in\mathbb{R}^{\frac{\bar{M}(\bar{M}+1)}{2}\times 1}$ contains the diagonal and lower-triangular entries of $\bar{\mathbf{B}}_{g,n}$.
$\mathbf{P}\in\{0,1\}^{\bar{M}^2\times\frac{\bar{M}(\bar{M}+1)}{2}}$ is the binary matrix mapping $\mathbf{b}_{g,n}$ into $\mathsf{vec}(\bar{\mathbf{B}}_{g,n})$, $\forall g\in\mathcal{G}$ defined as 
\begin{equation}
    \label{eq:p}
    [\mathbf{P}]_{\bar{M}(i-1)+i',l} = \begin{cases}
        1, & l = \frac{i(i-1)}{2}+i' ~\text{and}~ 1 \le i' \le i,\\
        1, & l = \frac{i'(i'-1)}{2}+i ~\text{and}~ i<i'\le \bar{M},\\
        0, & \text{otherwise},
    \end{cases}
\end{equation}
where $\forall i, i' \in\bar{\mathcal{M}}$.
According to (\ref{eq:wideband_constraint}), these vectors $\mathbf{b}_{g,n}$ for all subcarriers are jointly controlled by $\mathbf{b}_{\mathrm{c},g}\in\mathbb{R}^{\frac{\bar{M}(\bar{M}+1)}{2}\times 1}$ constructed with the entries $B_{\mathrm{c},m_g,m'_g}$ for $m'\ge m$.

Based on the above illustration, we can reformulate problem (\ref{eq:sub_prob1}) as the following form:
\begin{subequations}\label{eq:sub_prob2}
    \begin{align}
    \nonumber
    \max_{\mathbf{b}_{\mathrm{c},g}, \forall g} &\sum_{n\in\mathcal{N}}\Big|-y_{RT,n}+\sum_{g\in\mathcal{G}}\mathbf{y}_{RI,g,n}\\
    &\underbrace{~~\times(\jmath F_3(\overline{\mathsf{vec}}(\mathbf{P}\mathbf{b}_{g,n}))+Y_0\mathbf{I}_{\bar{M}})^{-1}\mathbf{y}_{IT,g,n}\Big|^2}_{=F(\mathbf{b}_{\mathrm{c}})}\\
    \label{eq:constraint_wideband}
    \text{s.t.} ~~&\mathbf{b}_{g,n} = F_1(\omega_n)\mathbf{b}_{\mathrm{c},g} + F_2(\omega_n), \forall n,g,\\
    \label{eq:constraint_range}
    &[\mathbf{b}_{\mathrm{c},g}]_l \in [B_\mathrm{min},B_\mathrm{max}], \forall g, l\in\mathcal{L},
    \end{align}
\end{subequations}
where $\mathbf{b}_{\mathrm{c}} = [\mathbf{b}_{\mathrm{c},1}^T,\ldots,\mathbf{b}_{\mathrm{c},G}^T]^T\in\mathbb{C}^{|\mathcal{L}|G\times 1}$, and $\mathcal{L} = \{1,\ldots,\frac{\bar{M}(\bar{M}+1)}{2}\}$.
The linear constraint (\ref{eq:constraint_range}) of each entry of $\mathbf{b}_{\mathrm{c},g}$ can be further removed by one-to-one mapping variables with arbitrary real values into the range $[B_\mathrm{min},B_\mathrm{max}]$. This can be done by introducing a tractable mapping function $[\mathbf{b}_{\mathrm{c},g}]_l = F_4([\mathbf{x}_g]_l)$, $\mathbf{x}_g\in\mathbb{R}^{\frac{\bar{M}(\bar{M}+1)}{2}\times 1}$, $\forall l\in\mathcal{L}$, such that 
\begin{enumerate}[$\bullet$]
    \item $F_4([\mathbf{x}_g]_l)$ is continuous, differentiable, and monotonously increasing in the full range $[\mathbf{x}_g]_l\in(-\infty,+\infty)$;
    \item $\lim_{[\mathbf{x}_g]_l\rightarrow +\infty}F_4([\mathbf{x}_g]_l) = B_\mathrm{max}$;
    \item $\lim_{[\mathbf{x}_g]_l\rightarrow -\infty}F_4([\mathbf{x}_g]_l) = B_\mathrm{min}$.
\end{enumerate}
An example of such a mapping function is given by \cite{nerini2024localized}:
\begin{equation}
    [\mathbf{b}_{\mathrm{c},g}]_l = F_4([\mathbf{x}_g]_l) = \frac{[\mathbf{x}_g]_l}{\sqrt{[\mathbf{x}_g]_l^2B_-^{-2} + 1}} + B_+, \forall g, l,
    \label{eq:mapping}
\end{equation}
where $B_- = \frac{B_\mathrm{max}-B_\mathrm{min}}{2}$ and $B_+ = \frac{B_\mathrm{max}+B_\mathrm{min}}{2}$. 
By this mapping function, we can transform problem (\ref{eq:sub_prob2}) into the following unconstrained optimization: 
\begin{equation}\label{eq:sub_prob3}
    \begin{aligned}
        \{\mathbf{x}_g^\star\}_{\forall g} = &\arg~\max_{\text{(\ref{eq:constraint_wideband}), (\ref{eq:mapping})}}~ F(\mathbf{b}_{\mathrm{c}}).
    \end{aligned}
\end{equation}
This can be directly solved by the quasi-Newton method, an iterative numerical method to find the local optimum based on the derivative of the objective function \cite{wright2006numerical}. In each iteration, a complexity $\mathcal{O}(\frac{G^2\bar{M}^2(\bar{M}+1)^2}{4})$ is required.

\subsubsection{Forest-Connected BD-RIS}
The proposed optimization method for group-connected BD-RIS with continuous-value admittance components can be easily adapted to forest-connected BD-RIS. 
This can be easily done by modifying the mapping $\bar{\mathbf{B}}_{g,n} = \overline{\mathsf{vec}}(\mathbf{P}\mathbf{b}_{g,n})$ for group-connected architecture based on (\ref{eq:symmetric}), (\ref{eq:B_cal}), and (\ref{eq:wideband_forest}) to align with the forest-connected architecture. More details are explained below.

For forest-connected BD-RIS, (\ref{eq:symmetric}), (\ref{eq:B_cal}), and (\ref{eq:wideband_forest}) imply that each matrix $\bar{\mathbf{B}}_{g,n}$ with entries $[\bar{\mathbf{B}}_{g,n}]_{m,m'} = B_{m_g,m'_g,n}$ is tridiagonal and symmetric. That is, $\bar{\mathbf{B}}_{g,n}$ is determined by its $\bar{M}$ diagonal and $\bar{M}-1$ non-zero lower/upper diagonal entries. 
This motivates us to extract these $2\bar{M} - 1$ entries from each $\bar{\mathbf{B}}_{g,n}$ to construct the vector $\breve{\mathbf{b}}_{g,n}\in\mathbb{R}^{(2\bar{M}-1)\times 1}$.
For ease of illustration, we define the mapping $\bar{\mathbf{B}}_{g,n} = F_5(\breve{\mathbf{b}}_{g,n})$, where $F_5(\cdot)$ is given such that 
\begin{equation}
    [\bar{\mathbf{B}}_{g,n}]_{m,m'} = \begin{cases}
        [\breve{\mathbf{b}}_{g,n}]_m, & m = m'\\
        [\breve{\mathbf{b}}_{g,n}]_{\bar{M}+m}, & m' = m + 1\\
        [\breve{\mathbf{b}}_{g,n}]_{\bar{M}+m'}, & m = m' + 1\\
        0, &\text{otherwise}
    \end{cases}.
\end{equation}
According to (\ref{eq:wideband_forest}), these vectors $\breve{\mathbf{b}}_{g,n}$ for all subcarriers are jointly controlled by $\breve{\mathbf{b}}_{\mathrm{c},g}\in\mathbb{R}^{(2\bar{M}-1)\times 1}$ constructed with the entries $B_{\mathrm{c},m_g,m_g}$ and $B_{\mathrm{c},m_g,{m+1}_g}$, $\forall m\in\bar{\mathcal{M}}$. Thus, we can transform problem (\ref{eq:sub_prob1}) into the following form:
\begin{subequations}\label{eq:sub_prob4}
    \begin{align}
    \nonumber
    \max_{\breve{\mathbf{b}}_{\mathrm{c},g}, \forall g} &\sum_{n\in\mathcal{N}}\Big|-y_{RT,n}+\sum_{g\in\mathcal{G}}\mathbf{y}_{RI,g,n}\\
    &\underbrace{~~\times(\jmath F_3(F_5(\breve{\mathbf{b}}_{g,n}))+Y_0\mathbf{I}_{\bar{M}})^{-1}\mathbf{y}_{IT,g,n}\Big|^2}_{=\breve{F}(\breve{\mathbf{b}}_{\mathrm{c}})}\\
    \label{eq:constraint_wideband_forest}
    \text{s.t.} ~~&\breve{\mathbf{b}}_{g,n} = F_1(\omega_n)\breve{\mathbf{b}}_{\mathrm{c},g} + F_2(\omega_n), \forall n,g,\\
    \label{eq:constraint_range_forest}
    &[\breve{\mathbf{b}}_{\mathrm{c},g}]_l \in [B_\mathrm{min},B_\mathrm{max}], \forall g, l\in\breve{\mathcal{L}},
    \end{align}
\end{subequations}
where $\breve{\mathbf{b}}_{\mathrm{c}} = [\breve{\mathbf{b}}_{\mathrm{c},1}^T,\ldots,\breve{\mathbf{b}}_{\mathrm{c},G}^T]^T\in\mathbb{C}^{|\breve{\mathcal{L}}|G\times 1}$, and $\breve{\mathcal{L}} = \{1,\ldots,2\bar{M}-1\}$. 
The constraint (\ref{eq:constraint_range_forest}) can be further eliminated by the mapping $[\breve{\mathbf{b}}_{\mathrm{c},g}]_l = F_4([\breve{\mathbf{x}}_g]_l)$ with $\breve{\mathbf{x}}_g\in\mathbb{R}^{(2\bar{M}-1)\times 1}$, $\forall l\in\breve{\mathcal{L}}$, such that problem (\ref{eq:sub_prob4}) can be transformed into an unconstrained optimization
\begin{equation}\label{eq:sub_prob5}
    \begin{aligned}
        \{\breve{\mathbf{x}}_g^\star\}_{\forall g} = &\mathop{\mathrm{arg~~~max}}\limits_{\text{(\ref{eq:constraint_wideband_forest})}, [\breve{\mathbf{b}}_{\mathrm{c},g}]_l = F_4([\breve{\mathbf{x}}_g]_l)}~\breve{F}(\breve{\mathbf{b}}_{\mathrm{c}}).
    \end{aligned}
\end{equation}
This can be solved by the quasi-Newton method with each iteration requiring complexity $\mathcal{O}(G^2(2\bar{M}-1)^2)$ \cite{wright2006numerical}.

\subsubsection{Summary}
With the solutions $\mathbf{x}_{g}^\star$, $\forall g\in\mathcal{G}$ to problem (\ref{eq:sub_prob3}) or $\breve{\mathbf{x}}_g^\star$, $\forall g\in\mathcal{G}$ to problem (\ref{eq:sub_prob5}), we can obtain 
\begin{enumerate}[$\bullet$]
    \item $\mathbf{b}_{\mathrm{c},g}^\star$ or $\breve{\mathbf{b}}_{\mathrm{c},g}^\star$ by (\ref{eq:mapping});
    \item $\mathbf{b}_{g,n}^\star$ by (\ref{eq:constraint_wideband}) or $\breve{\mathbf{b}}_{g,n}^\star$ by (\ref{eq:constraint_wideband_forest});
    \item $\mathbf{B}_{g,n}^\star = F_3(\overline{\mathsf{vec}}(\mathbf{P}\mathbf{b}_{g,n}^\star))$ or $\mathbf{B}_{g,n}^\star = F_3(F_5(\breve{\mathbf{b}}_{g,n}^\star))$, 
\end{enumerate} 
and $\mathbf{Y}_n^\star = \jmath\mathsf{blkdiag}(\mathbf{B}_{1,n}^\star,\ldots,\mathbf{B}_{G,n}^\star)$, $\forall n\in\mathcal{N}$, such that the effective channel $h_n^\star$ can be obtained by (\ref{eq:channel}).

\subsection{Solution to Problem (\ref{eq:sub_prob1}): Discrete-Value Admittance}
In this case, a straightforward solution is to apply the exhaustive search among all possible values of each tunable admittance. However, applying exhaustive search to solve the channel gain maximization problem requires $\mathcal{O}(2^{b\frac{G\bar{M}(\bar{M}+1)}{2}}G\bar{M}^3N)$ and $\mathcal{O}(2^{bG(2\bar{M}-1)}G\bar{M}^3N)$ for group- and tree-connected architectures, respectively. Both grow significantly with resolution bits $b$, group size $\bar{M}$, and number of groups $G$.
Another strategy is to directly perform the quantization based on the continuous-value solution. However, this strategy will cause large performance loss due to ignoring the coupling between admittance components at the optimization stage.
To reduce the computational complexity while maintaining good performance, we propose to iteratively optimize the discrete-value tunable admittance components of BD-RIS using a greedy algorithm.
The detailed solutions for group- and forest-connected architectures are explained below.

\subsubsection{Group-Connected BD-RIS}
In this case, the sum channel gain maximization problem (\ref{eq:sub_prob2}) writes as 
\begin{equation}\label{eq:sub_prob6}
    \mathbf{b}_{\mathrm{c}}^\star = \mathop{\mathrm{arg~~~max}}\limits_{\text{(\ref{eq:constraint_wideband})},~ [\mathbf{b}_{\mathrm{c},g}]_l \in \mathcal{B}} ~~F(\mathbf{b}_{\mathrm{c}}).
\end{equation}
To efficiently solve the problem, we propose to
\begin{enumerate}[$\bullet$]
    \item uniformly block the variable vector $\mathbf{b}_\mathrm{c}$ into $U$ sub-vectors, i.e. $\mathbf{b}_\mathrm{c} =[\mathbf{b}_1^T,\ldots,\mathbf{b}_{U}^T]^T$, where $\mathbf{b}_u = [\mathbf{b}_{\mathrm{c}}]_{(u-1)\bar{U}+1:u\bar{U}}\in\mathbb{C}^{\bar{U}\times 1}$, $\bar{U} = \frac{|\mathcal{L}|G}{U}$, $\forall u\in\mathcal{U} = \{1,\ldots,U\}$;
    \item construct the codebook for each sub-vector $\mathbf{b}_u$ based on the original constraint $[\mathbf{b}_{\mathrm{c},g}]_l\in\mathcal{B}$, as 
    \begin{equation}
        \bar{\mathcal{B}} = \{\mathbf{x}\in\mathbb{C}^{\bar{U}\times 1}~|~[\mathbf{x}]_i\in\mathcal{B}, \forall i = 1,\ldots,\bar{U}\}, 
    \end{equation}
    with $|\bar{\mathcal{B}}| = 2^{b\bar{U}}$.
\end{enumerate}

Then, the greedy algorithm is explained as follows. 
We first define the initialization $\mathbf{b}_{\mathrm{c}}^0 = [(\mathbf{b}_1^0)^T,\ldots,(\mathbf{b}_{U}^0)^T]^T$, where each sub-vector $\mathbf{b}_u^0$ is randomly chosen from the codebook $\bar{\mathcal{B}}$.
In the successive optimization of $\mathbf{b}_{u}^o$, $\forall u\in\mathcal{U}$, where the superscript denotes the $o$-th iteration, we aim to maximize the objective function $F([(\mathbf{b}_{1}^{o})^T,\ldots,(\mathbf{b}_{u-1}^{o})^T,\mathbf{b}_{u}^T,(\mathbf{b}_{u+1}^{o-1})^T,\ldots,(\mathbf{b}_U^{o-1})^T]^T)$ with $\mathbf{b}_{u}$ being the only variable vector to be optimized and other entries fixed. 
Then we have the following sub-problem:
\begin{equation}\label{eq:sub_prob7}
    \begin{aligned}
    \mathbf{b}_u^o
    = \mathop{\mathrm{arg~~~max}}\limits_{\text{(\ref{eq:constraint_wideband})},~ \mathbf{b}_{u} \in \bar{\mathcal{B}}} ~&F([(\mathbf{b}_{1}^{o})^T,\ldots,(\mathbf{b}_{u-1}^{o})^T,\mathbf{b}_{u}^T,\\ &~~(\mathbf{b}_{u+1}^{o-1})^T,\ldots,(\mathbf{b}_U^{o-1})^T]^T).
    \end{aligned}
\end{equation}
This problem can be efficiently solved by the exhaustive search over the set $\bar{\mathcal{B}}$ with proper options of $\bar{U}$ and $b$, which requires complexity $\mathcal{O}(2^{b\bar{U}}G\bar{M}^3N)$. By iteratively solving problem (\ref{eq:sub_prob7}) until the convergence of the objective function is achieved, we finally obtain the solution $\mathbf{b}_{\mathrm{c}}^\star$ to problem (\ref{eq:sub_prob6}).

\subsubsection{Forest-Connected BD-RIS}
The proposed optimization for group-connected BD-RIS with discrete-value admittance components is also readily extended to forest-connected BD-RIS. 
In this case, the sum channel gain maximization problem (\ref{eq:sub_prob4}) writes as 
\begin{equation}\label{eq:sub_prob8}
    \breve{\mathbf{b}}_{\mathrm{c}}^\star = \mathop{\mathrm{arg~~~max}}\limits_{\text{(\ref{eq:constraint_wideband_forest})},~[\breve{\mathbf{b}}_{\mathrm{c},g}]_l\in\mathcal{B}} ~ \breve{F}(\breve{\mathbf{b}}_{\mathrm{c}}),
\end{equation}
where each $\breve{\mathbf{b}}_u = [\breve{\mathbf{b}}_\mathrm{c}]_{(u-1)\bar{U}+1:u\bar{U}}$, $\forall u\in\mathcal{U}$ is iteratively optimized by the exhaustive search over $\bar{\mathcal{B}}$ such that the $\breve{\mathbf{b}}_{\mathrm{c}}^\star$ is obtained by the greedy algorithm.

\subsubsection{Summary}
With the solutions $\mathbf{b}_{\mathrm{c}}^\star$ to problem (\ref{eq:sub_prob6}) or $\breve{\mathbf{b}}_{\mathrm{c}}^\star$ to problem (\ref{eq:sub_prob8}), we can obtain 
\begin{enumerate}[$\bullet$]
    \item $\mathbf{b}_{g,n}^\star$ by (\ref{eq:constraint_wideband}) or $\breve{\mathbf{b}}_{g,n}^\star$ by (\ref{eq:constraint_wideband_forest});
    \item $\mathbf{B}_{g,n}^\star = F_3(\overline{\mathsf{vec}}(\mathbf{P}\mathbf{b}_{g,n}^\star))$ or $\mathbf{B}_{g,n}^\star = F_3(F_5(\breve{\mathbf{b}}_{g,n}^\star))$, 
\end{enumerate}
and the corresponding effective channel $h_n^\star$, $\forall n\in\mathcal{N}$ by (\ref{eq:channel}).

\subsection{Discussion}

The proposed optimization framework for BD-RIS-aided SISO-OFDM with group or forest-connected architectures and continuous or discrete-value admittance components can be readily generalized to MIMO-OFDM scenarios. 
Following the two-stage framework in Section \ref{sc:Optimization_SISO}-B, one can first ignore the inter-stream and/or inter-user interference in MIMO-OFDM channels and maximize the sum channel gain to optimize the frequency-dependent admittance matrices of BD-RIS following the proposed methods. With the optimized BD-RIS admittance matrices and the effective channel, existing precoding/combining techniques, such as multiple eigenmode transmission, zero-forcing (ZF), and minimum mean square error (MMSE), can be further adopted to effectively manage the interference. 
In this work, we only provide the results for BD-RIS-aided SISO-OFDM to focus on the impact of wideband modeling of BD-RIS. More detailed algorithm realization and performance evaluation for BD-RIS-aided MIMO-OFDM will be left as an interesting future work. 

\section{Performance Evaluation}
\label{sc:Performance}

In this section, we evaluate the performance of the BD-RIS-aided SISO-OFDM system to show the impact of the wideband modeling for BD-RIS with different architectures.

\subsection{Simulation Setup}
In the considered SISO-OFDM system, we assume the center frequency as $f_\mathrm{c} = 2.4$ GHz, the bandwidth as $B = 300$ MHz, and the number of subcarriers as $N=64$ to align with the operating frequencies of the varactor diode SMV1231-079 as illustrated in Section III-B. 
Therefore, the carrier frequencies are given by $f_n = f_\mathrm{c} + (n-\frac{N+1}{2})\frac{B}{N}$, $\forall n\in\mathcal{N}$.
The OFDM channels from the transmitter to the receiver, from the transmitter to BD-RIS, and from BD-RIS to the receiver, are modeled with the maximum delay spread of $D_{RT} = D_{RI} = D_{IT} = 16$ taps using CSCG random variables.  
The length of the CP is set to be 16. 
The distance-dependent channel pathloss comes from \cite{wu2021intelligent,zheng2019intelligent} and is modeled as $\zeta_s = \zeta_0d_s^{-\varepsilon_s}$, $\forall s\in\{RT,RI,IT\}$, where $\zeta_0 = -30$ dB refers to the signal attenuation at reference distance 1 m, $d_s$ denotes the distance between transmitter/BD-RIS/receiver, and $\varepsilon_s$ denotes the pathloss exponent. In the following simulations, we set $d_{RT} = 33$ m, $d_{RI} = 5$ m, $d_{IT} = 30$ m, $\varepsilon_{RT} = 3.8$, $\varepsilon_{RI} = 2.2$, and $\varepsilon_{IT} = 2.5$, which are typically assumed in existing RIS literature \cite{wu2021intelligent,zheng2019intelligent}.
The noise power is set as $\sigma^2 = -80$ dBm.

\begin{figure*}
    \centering
    \subfigure[Group-connected BD-RIS with and without considering wideband modeling ($M=36$)]{
    \includegraphics[width=0.32\textwidth]{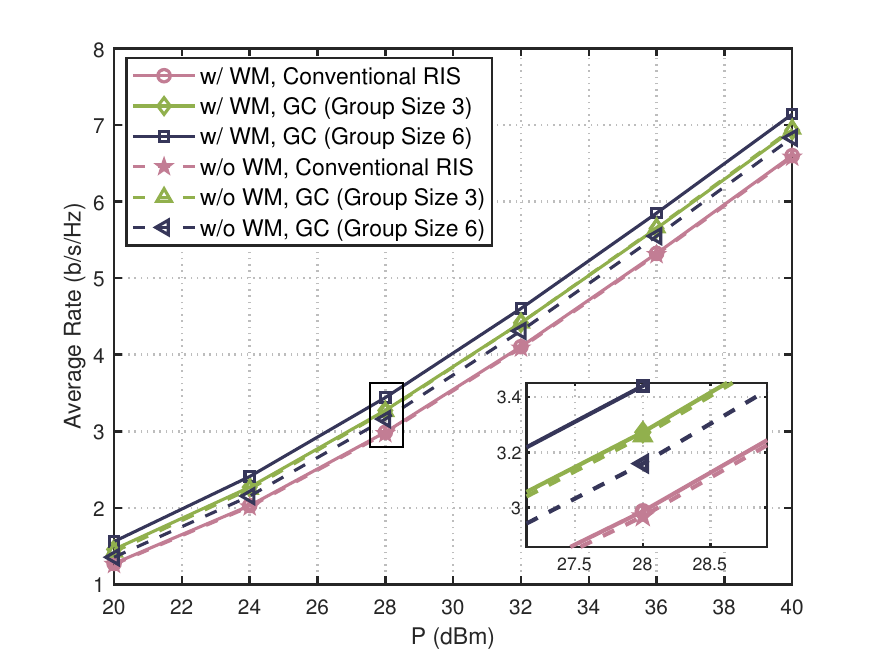}}
    \subfigure[Forest-connected BD-RIS with and without considering wideband modeling ($M=36$)]{
    \includegraphics[width=0.32\textwidth]{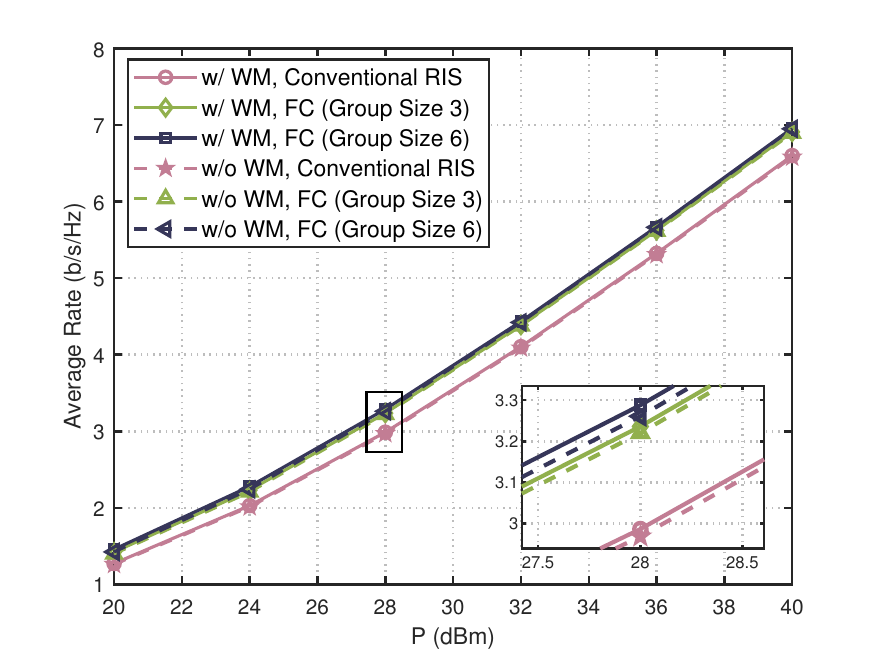}}
    \subfigure[Group- and forest-connected BD-RIS with wideband modeling ($M=36$)]{
    \includegraphics[width=0.32\textwidth]{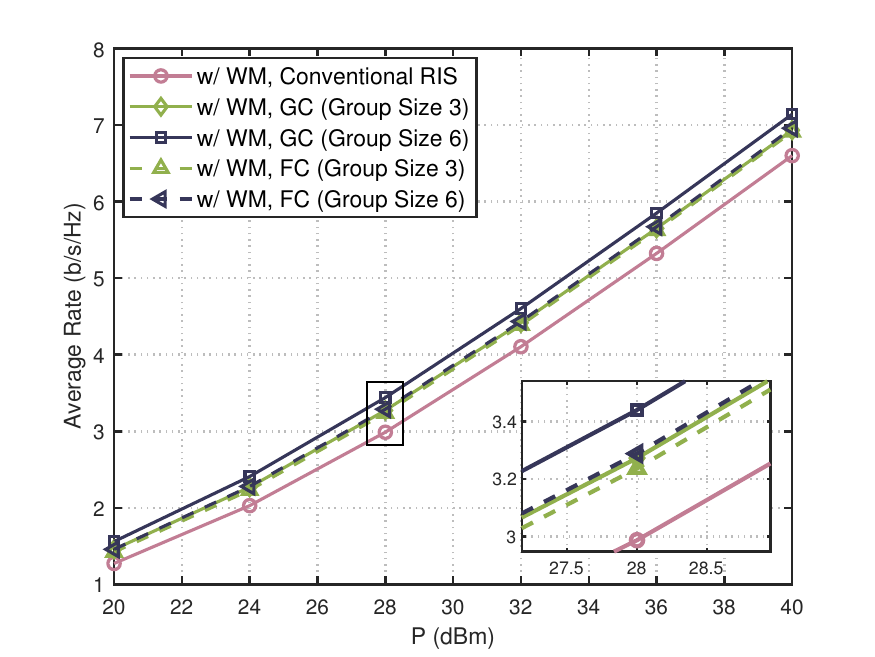}}
    \subfigure[Group-connected BD-RIS with and without considering wideband modeling ($M=48$)]{
    \includegraphics[width=0.32\textwidth]{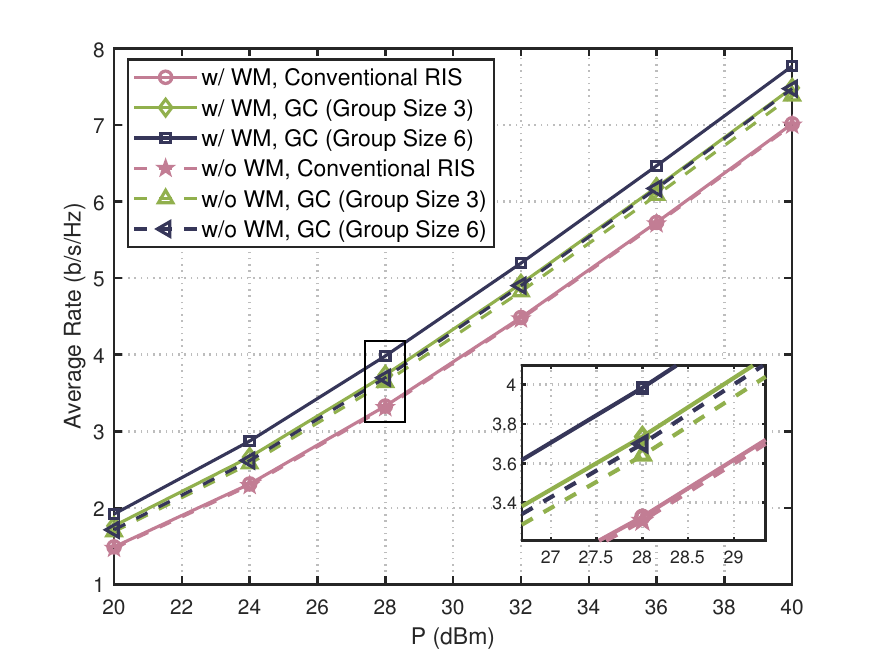}}
    \subfigure[Forest-connected BD-RIS with and without considering wideband modeling ($M=48$)]{
    \includegraphics[width=0.32\textwidth]{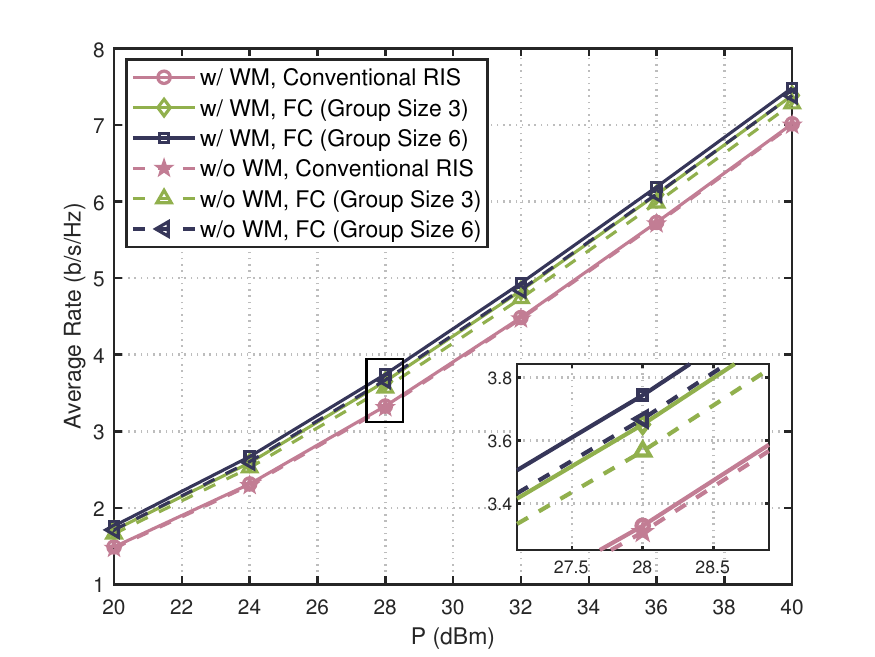}}
    \subfigure[Group- and forest-connected BD-RIS with wideband modeling ($M=48$)]{
    \includegraphics[width=0.32\textwidth]{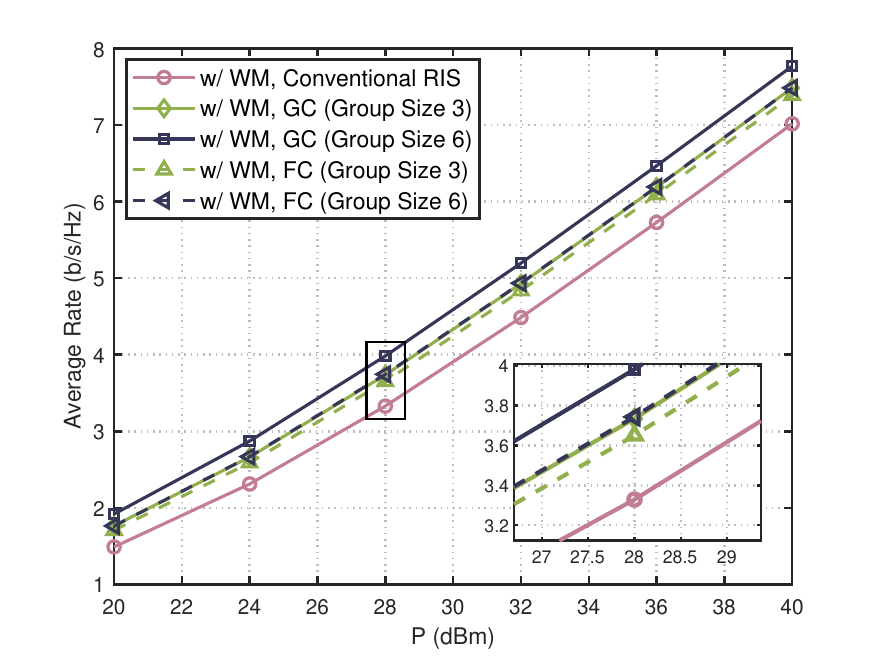}}
    \caption{Average rate versus transmit power $P$ with BD-RIS having different architectures ($M\in\{36,48\}$, $\bar{M}\in\{1,3,6\}$). The legend ``WM'' is short for wideband modeling; ``GC'' is short for group-connected; ``FC'' is short for forest-connected. For both ``GC'' and ``FC'' architectures, the case of $\bar{M}=1$ refers to the conventional RIS.}
    \label{fig:AR_P}
\end{figure*}

\subsection{Benchmark Schemes}
\subsubsection{Frequency-Independent BD-RIS} 
To show how the wideband modeling of BD-RIS impacts the system performance, we include the case where the BD-RIS admittance matrices are designed such that they remain the same for all subcarriers. 
That is, we assume $\mathbf{Y}_n = \mathbf{Y}$, $\forall n\in\mathcal{N}$ as in \cite{soleymani2024maximizing,demir2024wideband} where the narrowband BD-RIS model is adopted in wideband scenarios. In other words, we have 
\begin{equation}
    \bar{h}_n = \frac{1}{2Y_0}(-y_{RT,n} + \mathbf{y}_{RI,n}(\mathbf{Y} + Y_0\mathbf{I}_M)^{-1}\mathbf{y}_{IT,n}), \forall n.
\end{equation}  
Then the proposed algorithms for either continuous or discrete-value admittance components can also be adopted to solve the following problem:
\begin{equation}
    \begin{aligned}
    &\{\{Y_{m_g,m'_g}^\sharp\}_{\forall m,m',g}, \{p_n^\sharp\}_{\forall n}\}\\
    &~~ = \mathop{\mathrm{arg~~~max}}\limits_{\substack{\text{(\ref{eq:Y_blkdiag})-(\ref{eq:Y_forest})}, \sum_{n\in\mathcal{N}} p_n\le P\\ \Im\{Y_{m_g,m'_g}\}\in[B_\mathrm{min},B_\mathrm{max}]}} ~\frac{1}{N}\sum_{n\in\mathcal{N}}\log_2\Big(1+\frac{p_n|\bar{h}_n|^2}{\sigma^2}\Big),
    \end{aligned}
\end{equation} 
where constraints (\ref{eq:Y_blkdiag})-(\ref{eq:Y_calculate}) refer to the group-connected architecture and (\ref{eq:Y_blkdiag})-(\ref{eq:Y_forest}) refer to the forest-connected architecture. 
The obtained solutions $Y_{m_g,m'_g}^\sharp$, $\forall m,m'\in\bar{\mathcal{M}}$, $\forall g\in\mathcal{G}$ are further plugged into (\ref{eq:wideband_constraint}) or (\ref{eq:wideband_forest}), such that we can obtain 
\begin{enumerate}[$\bullet$]
    \item the admittance matrices $\mathbf{Y}_n^\sharp$, $\forall n\in\mathcal{N}$ by (\ref{eq:Y_blkdiag})-(\ref{eq:Y_forest});
    \item the channels $h_n^\sharp$, $\forall n\in\mathcal{N}$ by (\ref{eq:channel}),
\end{enumerate} 
together with $p_n^\sharp$, $\forall n\in\mathcal{N}$ to be plugged into the original objective function (\ref{eq:prob0}) for fair performance comparisons.

\subsubsection{Conventional RIS}
To show how the performance of BD-RIS compares to that of conventional RIS in the presence or absence of wideband modeling, we include the case where the BD-RIS has diagonal admittance matrices. In other words, the conventional RIS is a special case of the group or forest-connected BD-RIS with $\bar{M}=1$ such that the proposed methods can be directly implemented.

\begin{figure*}
    \centering
    \subfigure[Group-connected BD-RIS with discrete-value admittance components ($M=36$)]{
    \includegraphics[width=0.32\textwidth]{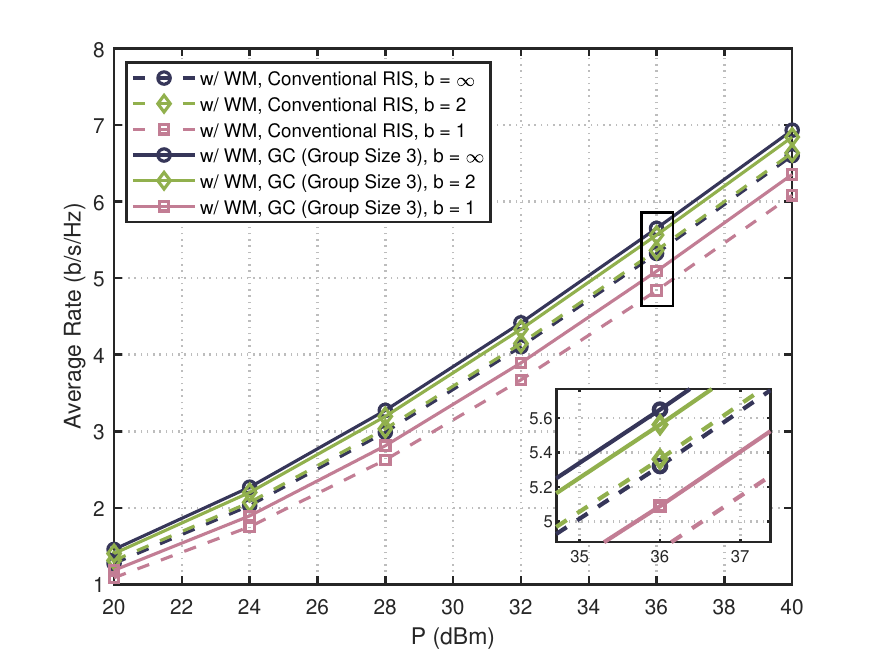}}
    \subfigure[Forest-connected BD-RIS with discrete-value admittance components ($M=36$)]{
    \includegraphics[width=0.32\textwidth]{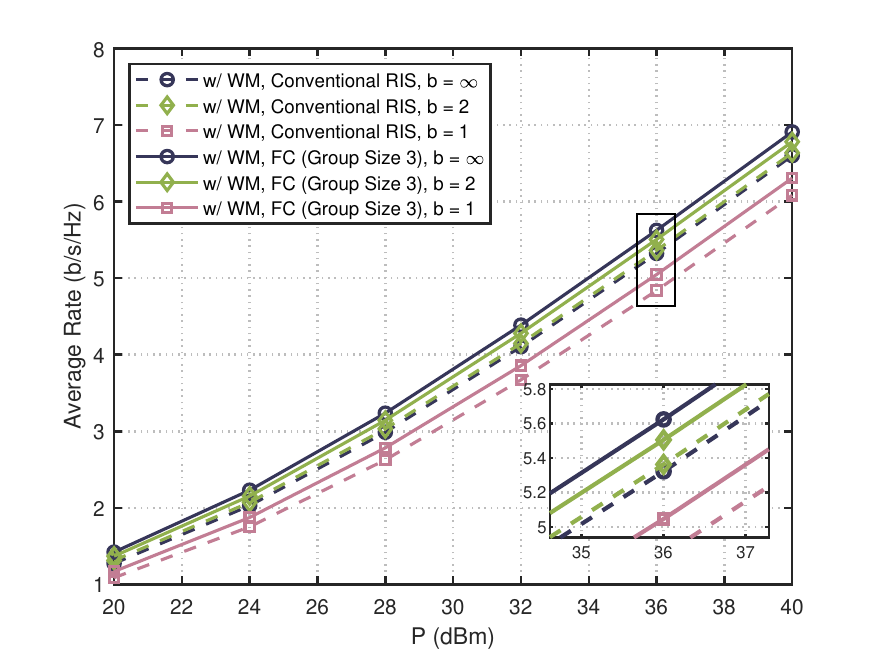}}
    \subfigure[BD-RIS with and without considering wideband modeling ($M=36$, $\bar{M}=3$)]{
    \includegraphics[width=0.32\textwidth]{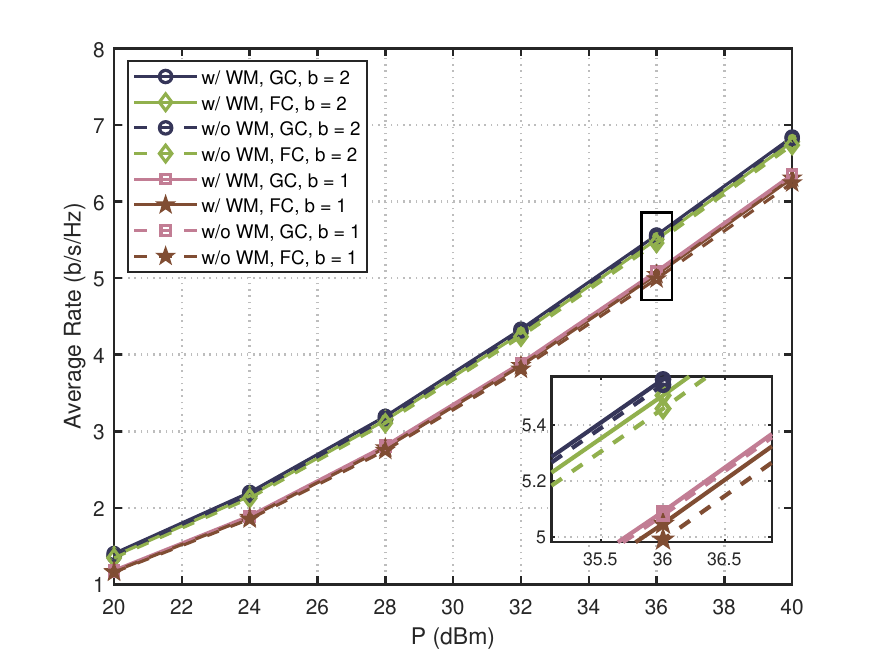}}
    \subfigure[Group-connected BD-RIS with discrete-value admittance components ($M=48$)]{
    \includegraphics[width=0.32\textwidth]{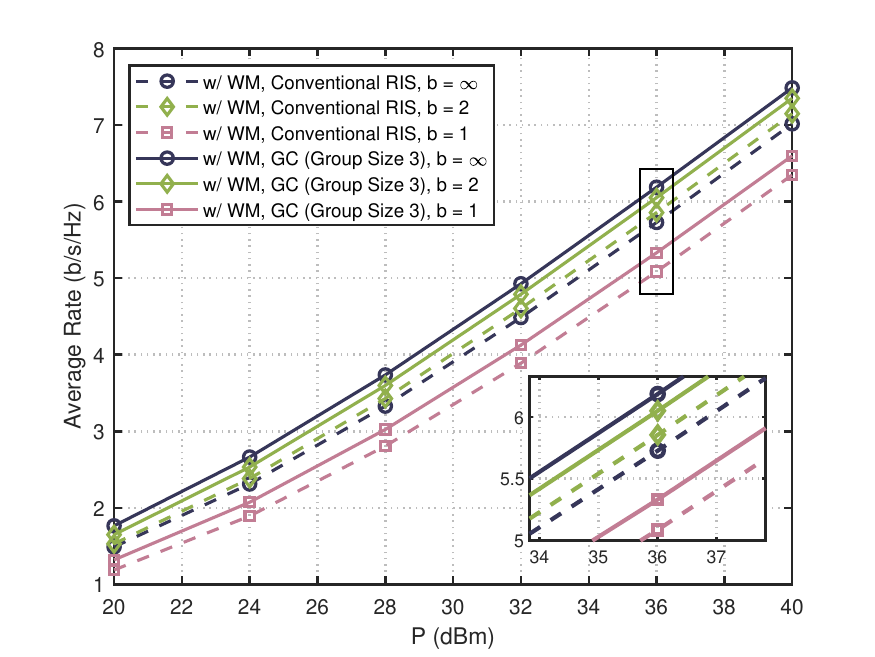}}
    \subfigure[Forest-connected BD-RIS with discrete-value admittance components ($M=48$)]{
    \includegraphics[width=0.32\textwidth]{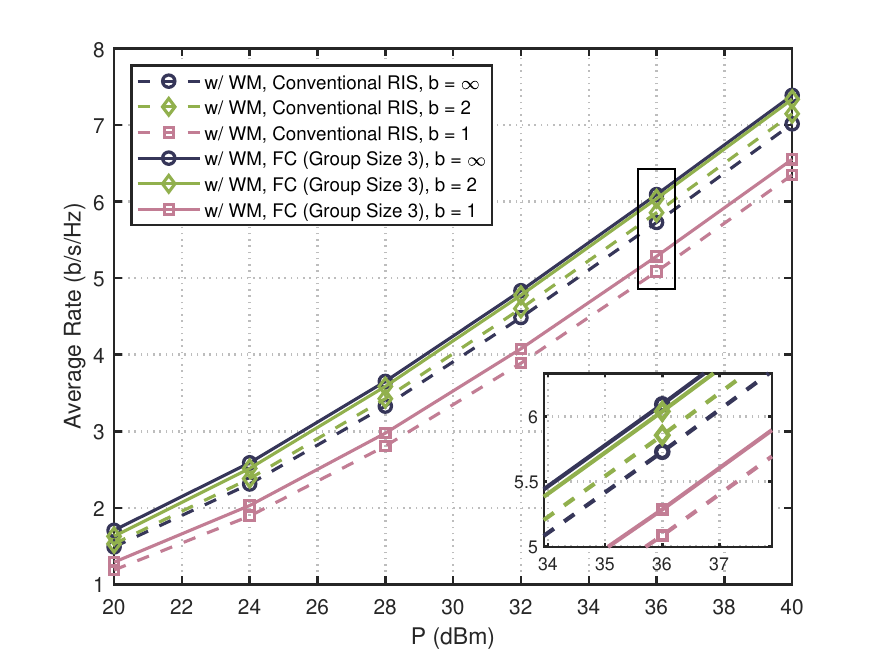}}
    \subfigure[BD-RIS with and without considering wideband modeling ($M=48$, $\bar{M}=3$)]{
    \includegraphics[width=0.32\textwidth]{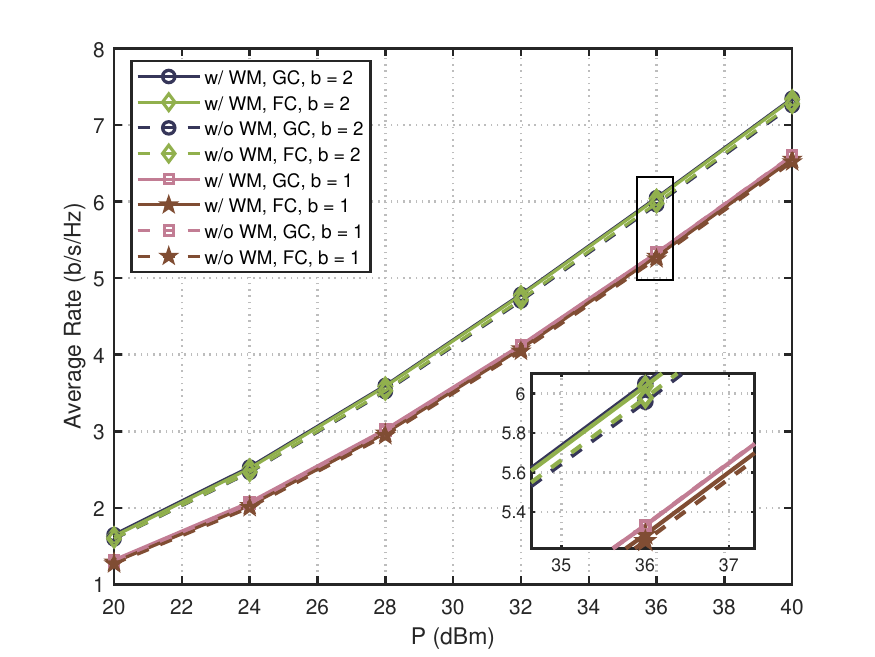}}   
    \caption{Average rate versus transmit power $P$ with BD-RIS having different architectures and discrete-value admittance components ($M\in\{36,48\}$, $\bar{M}\in\{1,3\}$, $b\in\{1,2\}$. $\bar{U}=4$ for $b=1$; $\bar{U}=2$ for $b=2$). The legend ``WM'' is short for wideband modeling; ``GC'' is short for group-connected; ``FC'' is short for forest-connected. For both ``GC'' and ``FC'' architectures, the case of $\bar{M}=1$ refers to the conventional RIS.}
    \label{fig:AR_P_discrete}
\end{figure*}

\subsection{System Performance}

\subsubsection{Continuous-Value Admittance}
We evaluate the performance of the proposed wideband modeling and algorithm design by plotting 
the average rate of the BD-RIS-aided SISO-OFDM system versus the transmit power $P$ in Fig. \ref{fig:AR_P}, from which we have the following observations. 

\textit{First}, when taking into account the wideband modeling, BD-RIS with larger group size of the group-connected architecture achieves better performance within all the considered range of transmit power. 
This is attributed to the increasing wave manipulation capabilities offered by more complex circuit topologies of BD-RIS. 
There is an increasing performance gap between taking into account the wideband effect for designing group-connected BD-RIS or not. For example, the performance gap is around 5\% with group size 3, while it reaches around 8\% with group size 6. 
More importantly, when ignoring the wideband effect for designing group-connected BD-RIS, BD-RIS architectures with larger group size cannot always provide better performance. 
For example, BD-RIS with group size $\bar{M}=3$ outperforms that with $\bar{M} = 6$ when $M = 36$.
This is because the variation for BD-RIS matrices between different subcarriers becomes more significant with increasing number of admittances, such that directly designing BD-RIS ignoring this variation will lead to increasing performance loss.

\textit{Second}, for the cases of both with or without wideband modeling, BD-RIS with larger group size of the forest-connected architecture achieves better performance within all the considered range of transmit power, while the performance enhancement is not as significant as the group-connected architecture due to the reduced number of tunable admittance components.  
This also demonstrates that the impact of wideband modeling for BD-RIS to the system performance is proportional to the circuit complexity of BD-RIS architectures: The higher the circuit complexity of BD-RIS architectures, the larger the performance gap between schemes with and without including wideband modeling of BD-RIS.  

\textit{Third}, when taking into consideration the wideband modeling of BD-RIS, BD-RIS having group-connected architectures outperform that having forest-connected architectures with the same group size. While group- and forest-connected architectures achieve the same performance in narrowband point-to-point systems \cite{nerini2023beyond}, wideband OFDM systems benefit from more tunable admittance components in BD-RIS to deal with the frequency selectivity.
This also suggests that the observations and pareto frontier in \cite{nerini2023beyond} and \cite{nerini2023pareto} will change for wideband compared to narrowband.

\subsubsection{Discrete-Value Admittance}
In Fig. \ref{fig:AR_P_discrete}, we plot the average rate of the BD-RIS-aided SISO-OFDM system versus the transmit power when the BD-RIS has discrete-value admittance components controlled by $b$ bits\footnote{According to our simulations, the value of $\bar{U}$ does not impact the performance for conventional RIS, and a larger $\bar{U}$ will lead to slightly increasing average rate for BD-RIS architectures. Therefore, to save on the complexity without sacrificing too much the performance, we set $\bar{U}=4$ for $b=1$ and $\bar{U}=2$ for $b=2$.}. From Fig. \ref{fig:AR_P_discrete}, we have the following observations. 

\textit{First}, when taking into account the wideband modeling, for both group-connected and forest-connected BD-RIS, increasing the number of resolution bits for each discrete-value admittance leads to better performance. In addition, BD-RIS with 2-bit resolution for each discrete-value admittance can achieve satisfactory performance close to that with continuous-value admittance components. 

\textit{Second}, similar to the continuous-value admittance cases, in discrete-value cases, there also exists performance gap between including or ignoring the wideband modeling of BD-RIS in the optimization stage. More importantly, with relatively small numbers of resolution bits $b$ and small group size $\bar{M}$ of BD-RIS architectures, the performance loss coming from ignoring the wideband modeling is negligible. 
This observation indicates that it is efficient to directly adopt the narrowband BD-RIS model in wideband scenarios when using simple BD-RIS architectures with limited freedom in tuning admittance components.

\section{Conclusion}
\label{sc:Conclusion}

In this work, we study the wideband modeling and optimization of BD-RIS in a SISO-OFDM system. Particularly, we first derive the system and signal model for BD-RIS-aided OFDM systems, which bridges the time and frequency-domain channels and shows explicitly the frequency dependency of the scattering matrix for BD-RIS. We next characterize this frequency dependency by modeling each tunable admittance of BD-RIS based on lumped circuits. Taking into account the practical frequency bands, we propose a simple linear model with respect to frequency. 

With the proposed BD-RIS model and BD-RIS-aided signal model, we consider the optimization of a BD-RIS-aided SISO-OFDM system with the aim of maximizing the average rate. For BD-RIS with continuous-value admittance components, we propose to transform the original problem into an unconstrained optimization problem to be solved by the well-known quasi-Newton method. For BD-RIS with discrete-value admittance components, we adopt the idea of the greedy algorithm to successively design each tunable admittance. 

Finally, we provide simulation results to evaluate the effectiveness of the proposed algorithms and the impact of the wideband modeling. 
Results show that BD-RIS always outperforms conventional RIS whether the wideband modeling is taken into account during the optimization process or not.
In addition, the higher the circuit complexity of BD-RIS, the larger the impact of the wideband modeling on the system performance. This demonstrates the importance of accurately capturing the frequency-dependent response of BD-RIS with complex architectures in wideband communication systems.
Future research avenues include but are not limited to 1) designing faster optimization methods, such as designing a codebook scheme for BD-RIS aided scenarios, as in conventional RIS cases \cite{an2022codebook}, to ease the requirement to accurate channel state information, and 2) investigating other performance metrics beyond rate and developing beamforming design algorithms to show the impact of wideband modeling at BD-RIS from multiple perspectives.

\begin{appendix}[Proof of Proposition 1]
    We first rewrite the frequency-domain channel $\mathbf{H}$ by re-arranging the block-circulant matrices. This yields
    \begin{equation}
        \begin{aligned}
            \mathbf{H} = &\mathbf{F}_N(\bar{\mathbf{H}}_{RT} + \bar{\mathbf{H}}_{RI}\bar{\mathbf{\Theta}}\bar{\mathbf{H}}_{IT})\mathbf{F}_N^H\\
            \overset{\text{(a)}}{=} &\mathbf{F}_N(\bar{\mathbf{H}}_{RT} + \bar{\mathbf{H}}_{RI}\mathbf{\Gamma}\mathbf{\Gamma}^T\bar{\mathbf{\Theta}}\mathbf{\Gamma}\mathbf{\Gamma}^T\bar{\mathbf{H}}_{IT})\mathbf{F}_N^H,
            % \nonumber
            % \overset{\text{(b)}}{=} &\mathbf{F}_N\bar{\mathbf{H}}_{RT}\mathbf{F}_N^H + \mathbf{F}_N[\bar{\mathbf{H}}_{RI,1},\ldots,\bar{\mathbf{H}}_{RI,M}]\\
            % &~~~~~\times \begin{bmatrix}
            %     \bar{\mathbf{\Theta}}_{1,1} &\cdots &\bar{\mathbf{\Theta}}_{1,M}\\
            %     \vdots & \ddots & \vdots\\
            %     \bar{\mathbf{\Theta}}_{M,1} &\cdots &\bar{\mathbf{\Theta}}_{M,M}
            % \end{bmatrix}
            % \begin{bmatrix}
            %     \bar{\mathbf{H}}_{IT,1}\\
            %     \vdots \\
            %     \bar{\mathbf{H}}_{IT,M}
            % \end{bmatrix}\mathbf{F}_N^H,
        \end{aligned}
    \end{equation}
    where (a) holds by introducing the permutation matrix $\mathbf{\Gamma}\in\{0,1\}^{NM\times NM}$ with $\mathbf{\Gamma}\mathbf{\Gamma}^T=\mathbf{I}_{NM}$, which transforms a block-circulant matrix into a row-wise concatenation of circulant matrices \cite{kwon2017hybrid}. Specifically, we have  
    \begin{equation}
        [\mathbf{\Gamma}]_{i,j} = \begin{cases}
            1, &i = (n-1)M+m, j = (m-1)N+n\\
            0, &\text{otherwise}
        \end{cases},
    \end{equation}
    with $\forall n\in\mathcal{N}$ and $\forall m\in\mathcal{M}$, such that
    \begin{equation}
        \begin{aligned} 
            &\bar{\mathbf{H}}_{RI}\mathbf{\Gamma} = [\bar{\mathbf{H}}_{RI,1},\ldots,\bar{\mathbf{H}}_{RI,M}],\\
            &\mathbf{\Gamma}^T\bar{\mathbf{H}}_{IT} = [\bar{\mathbf{H}}_{IT,1}^T,\ldots,\bar{\mathbf{H}}_{IT,M}^T]^T,\\
            &\mathbf{\Gamma}^T\bar{\mathbf{\Theta}}\mathbf{\Gamma} = \begin{bmatrix}
                \bar{\mathbf{\Theta}}_{1,1} &\cdots &\bar{\mathbf{\Theta}}_{1,M}\\
                \vdots & \ddots & \vdots\\
                \bar{\mathbf{\Theta}}_{M,1} &\cdots &\bar{\mathbf{\Theta}}_{M,M}
            \end{bmatrix},
        \end{aligned}
    \end{equation}
    where $\bar{\mathbf{H}}_{RI,i}\in\mathbb{C}^{N\times N}$, $\bar{\mathbf{H}}_{IT,j}\in\mathbb{C}^{N\times N}$, and $\bar{\mathbf{\Theta}}_{i,j}\in\mathbb{C}^{N\times N}$, $\forall i,j\in\mathcal{M}$, respectively, are circulant matrices with entries 
    \begin{equation}
        \begin{aligned}
            &[\bar{\mathbf{H}}_{RI,i}]_{:,p} = [\bar{\mathbf{H}}_{RI}]_{:,(p-1)M+i}, \forall i,\forall p\in\mathcal{N},\\
            &[\bar{\mathbf{H}}_{IT,j}]_{q,:} = [\bar{\mathbf{H}}_{IT}]_{(q-1)M+j,:}, \forall j,\forall q\in\mathcal{N},\\
            &[\bar{\mathbf{\Theta}}_{i,j}]_{p,q} =  [\bar{\mathbf{\Theta}}]_{(p-1)M+i,(q-1)M+j}, \forall i,j,p,q.
        \end{aligned}
    \end{equation}
    Then, we can rewrite $\mathbf{H}$ as
    \begin{equation}
        \begin{aligned} 
            \mathbf{H} = &\mathbf{F}_N\bar{\mathbf{H}}_{RT}\mathbf{F}_N^H\\
             &~~+ \mathbf{F}_N\sum_{i\in\mathcal{M}}\sum_{j\in\mathcal{M}} \bar{\mathbf{H}}_{RI,i}\bar{\mathbf{\Theta}}_{i,j}\bar{\mathbf{H}}_{IT,j}\mathbf{F}_N^H\\
             \overset{\text{(b)}}{=} &\mathbf{F}_N\bar{\mathbf{H}}_{RT}\mathbf{F}_N^H + \sum_{i\in\mathcal{M}}\sum_{j\in\mathcal{M}} \mathbf{F}_N\bar{\mathbf{H}}_{RI,i}\mathbf{F}_N^H\\
             &~~\times\mathbf{F}_N\bar{\mathbf{\Theta}}_{i,j}\mathbf{F}_N^H\mathbf{F}_N\bar{\mathbf{H}}_{IT,j}\mathbf{F}_N^H,
        \end{aligned}
    \end{equation}
    where (b) holds due to the property of the normalized DFT, i.e., $\mathbf{F}_N^H\mathbf{F}_N = \mathbf{I}_N$.
    Since a circulant matrix can be diagonalized by DFT \cite{simon2015operator}, we have that 
    \begin{equation}
        \begin{aligned}
            &\mathbf{F}_N\bar{\mathbf{H}}_{RT}\mathbf{F}_N^H = \mathsf{diag}(h_{RT,1},\ldots,h_{RT,N}),\\ 
            &\mathbf{F}_N\bar{\mathbf{H}}_{RI,i}\mathbf{F}_N^H = \mathsf{diag}(h_{RI,i,1},\ldots,h_{RI,i,N}),\\
            &\mathbf{F}_N\bar{\mathbf{H}}_{IT,j}\mathbf{F}_N^H = \mathsf{diag}(h_{IT,j,1},\ldots,h_{IT,j,N}),\\
            &\mathbf{F}_N\bar{\mathbf{\Theta}}_{i,j}\mathbf{F}_N^H = \mathsf{diag}(\theta_{i,j,1},\ldots,\theta_{i,j,N}),
        \end{aligned}
    \end{equation}
    where $h_{RT,n}\in\mathbb{C}$, $h_{RI,i,n}\in\mathbb{C}$, and $h_{IT,j,n}\in\mathbb{C}$, respectively, denote frequency-domain channels from the transmitter to receiver, from the $i$-th element of BD-RIS to the receiver, and from the transmitter to the $j$-th element of BD-RIS at subcarrier $n$, $\forall i,j\in\mathcal{M}$, $\forall n\in\mathcal{N}$. 
    $\theta_{i,j,n}\in\mathbb{C}$ denotes the scattering parameter from the $j$-th element to the $i$-th element of BD-RIS at subcarrier $n$. 
    By introducing the frequency-domain channels $\mathbf{h}_{RI,n}\in\mathbb{C}^{1\times M}$, $\mathbf{h}_{IT,n}\in\mathbb{C}^{M\times 1}$, and the scattering matrix of BD-RIS $\mathbf{\Theta}_n\in\mathbb{C}^{M\times M}$ at subcarrier $n$ with entries 
    \begin{equation}
        \begin{aligned}
            &[\mathbf{h}_{RI,n}]_i = h_{RI,i,n},~ [\mathbf{h}_{IT,n}]_j = h_{IT,j,n}, \\
            &[\mathbf{\Theta}_n]_{i,j} = \theta_{i,j,n}, ~\forall i,j,n,
        \end{aligned} 
    \end{equation}
    we finally have 
    \begin{equation}
        \begin{aligned}
            \mathbf{H} = &\mathsf{diag}(h_{RT,1} + \mathbf{h}_{RI,1}\mathbf{\Theta}_1\mathbf{h}_{IT,1},\\
            &~~\ldots,h_{RT,N} + \mathbf{h}_{RI,N}\mathbf{\Theta}_N\mathbf{h}_{IT,N}).
        \end{aligned}  
    \end{equation}
    The proof is completed. \hfill$\square$
\end{appendix}

\bibliographystyle{IEEEtran}
\bibliography{refs}

\end{document}